\documentclass[aip, amsmath, amssymb, reprint]{revtex4-1}

\usepackage{graphicx}
\usepackage{dcolumn}
\usepackage{bm}

\usepackage[utf8]{inputenc}
\usepackage[T1]{fontenc}
\usepackage{mathptmx}
\usepackage{etoolbox}
\usepackage{lipsum}
\usepackage[dvipsnames]{xcolor}
\usepackage{booktabs}
\usepackage{tikz}

\usepackage{styling}
\usepackage{tikz_commands}
\draft

\makeatletter
\def\@email#1#2{
 \endgroup
 \patchcmd{\titleblock@produce}
  {\frontmatter@RRAPformat}
  {\frontmatter@RRAPformat{\produce@RRAP{*#1\href{mailto:#2}{#2}}}\frontmatter@RRAPformat}
  {}{}
}
\makeatother

\newcommand{\LMU}{Arnold Sommerfeld Center for Theoretical Physics, Center for NanoScience, and Munich Center for Quantum Science and Technology, Ludwig-Maximilians-Universität München, 80333 Munich, Germany}

\newcommand{\code}[1]{\texttt{#1}}

\begin{document}

\title{KeldyshQFT: A C++ codebase for real-frequency multiloop functional renormalization
group and parquet computations of the single-impurity Anderson model}

\author{Nepomuk Ritz$^*$}
\thanks{These authors contributed equally to this work.}
\affiliation{\LMU}
\email[]{nepomuk.ritz@physik.uni-muenchen.de}

\author{Anxiang Ge}
\thanks{These authors contributed equally to this work.}
\affiliation{\LMU}

\author{Elias Walter}
\affiliation{\LMU}

\author{Santiago Aguirre}
\affiliation{\LMU}

\author{Jan von Delft}
\affiliation{\LMU}

\author{Fabian B.~Kugler}
\affiliation{Center for Computational Quantum Physics, Flatiron Institute, 162 5th Avenue, New York, NY 10010, USA}

\date{\today}

\begin{abstract}
We provide a detailed exposition of our computational
framework designed for the accurate calculation of real-frequency
dynamical correlation functions of the single-impurity Anderson model
(AM) in the regime of weak to intermediate coupling. Using quantum
field theory within the Keldysh formalism to directly access the
self-energy and dynamical susceptibilities in real frequencies, as detailed in our recent publication \cite{main_paper}, the primary computational challenge is the full
three-dimensional real-frequency dependence of the four-point vertex.
Our codebase provides a fully MPI+OpenMP parallelized implementation of the
functional renormalization group (fRG) and the self-consistent parquet equations within the parquet
approximation. It leverages vectorization to handle the additional
complexity imposed by the Keldysh formalism, using optimized data
structures and highly performant integration routines. Going beyond the
results shown in the previous publication, the code includes functionality to
perform fRG calculations in the multiloop framework, at arbitrary loop
order, including self-consistent self-energy iterations. Moreover,
implementations of various regulators, such as hybridization, interaction, frequency, and temperature are supplied.
\end{abstract}


\maketitle 

\tableofcontents

\section{Introduction}\label{sec:introduction}

In the study of strongly correlated electrons, dynamical correlation functions are quantities of major interest, as they provide insights into the collective behavior and emergent phenomena arising from electronic interactions. Capturing the effects of two-particle (or four-point) correlations is one of the current major frontiers in the field. Their dynamical properties are inherently difficult to compute, as they involve three independent frequency arguments.

While most previous works on this subject focused on four-point functions in imaginary frequencies in the Matsubara formalism\cite{10.1143/PTP.14.351, abrikosov2012methods} (MF), obtaining real-frequency information is crucial for direct comparisons to experiments. The extraction of real-frequency data from the results of a calculation in the MF is in principle possible via analytic continuation\cite{Baym1961}. However, it is hard to do so reliably in practice, as the conditions for the procedure outlined in \onlinecite{Baym1961} are not met by finite amounts of numerical data. This renders analytic continuation an ill-defined problem, despite numerous attempts \cite{JARRELL1996133, PhysRevB.57.10287, PhysRevB.93.075104}. Furthermore, it had not been worked out in full detail until very recently\cite{ge2023analytic}, how analytic continuation of four-point functions could be achieved even under the assumption of analytically available results.

Pioneering attempts to directly compute real-frequency dynamical four-point correlation functions using simplified approaches made use of diagrammatic ladder approximations \cite{Kroha1997, Kroha1998} or were restricted to a simplified frequency dependence\cite{jakobs_functional_2010, jakobs_nonequilibrium_2010, jakobs_properties_2010}. The first fully unbiased treatment of the fluctuations contributing to the four-point vertex was achieved only a few years ago using a multipoint extension of the numerical renormalization group (NRG) \cite{Kugler2021, Lee2021}. 

Even more recently, we presented a similarly unbiased treatment of the four-point vertex of the single-impurity Anderson model using a QFT framework within the Keldysh formalism (KF), employing two related diagrammatic methods: the functional renormalization group (fRG) and the self-consistent parquet equations in the parquet approximation \cite{main_paper}. While we focused on the conceptual aspects and discussed the performance of the methods in great detail in the previous publication, here we wish to provide a detailed exposition of the computational framework for the numerical calculations of self-energies and vertex functions. In addition to what was shown in Ref.\,\onlinecite{main_paper}, the code discussed in this paper is capable of performing fRG calculations in the multiloop framework up to arbitrary loop order, which connects the fRG to the parquet formalism \cite{kugler_multiloop_2018, kugler_multiloop_2018-1, kugler_derivation_2018}.

This paper aims to serve as a reference for future extensions or revisions of the code. The codebase discussed here was developed by several people over the course of multiple years, during which some goals and priorities changed and the code had to be adapted accordingly. This paper shall document how the code works and what was learned during its development. 

Some general design choices during development resulted in convenient features of the code and are recommended for future projects. In the following, we briefly discuss the most important features:

\paragraph{Modularity.} Every main building block of the code and each functionality is implemented individually, using classes and functions that serve one purpose only. As a consequence, a developer can keep an overview of the functionality. It is also comparatively easy to reuse existing features and combine them into new functionality. For example, for both the computation of the Schwinger-Dyson equation during parquet computations and the evaluation of the flow equation for the self-energy during the solution of an mfRG flow, the same classes for vertices, propagators, self-energies, and the same function for contracting a loop are used, as described in Sec.\,\ref{sec:objects} and Sec.\,\ref{sec:functionality}. In addition, modularity enables unit-testing of each functionality, something too often ignored during research software development. Modularity is probably the most important feature that should be prioritized in developing any research software.

\paragraph{Flexibility.} A modular design makes the code flexible, too. Some additional choices were made to improve its flexibility even further. Most importantly, the code enables computations in three different formalisms: The finite-temperature Matsubara formalism (MF), the zero-temperature Matsubara formalism, and the Keldysh formalism (KF), which works at any temperature and generalizes to systems out of thermal equilibrium. Consequently, some functionality had to be implemented multiple times, such as contractions, which require summations over discrete Matsubara frequencies in the finite-temperature MF but integrations over continuous frequencies in the zero-temperature MF and the KF. Additionally, in the KF, all quantities are complex-valued, whereas they are real-valued in the MF for particle-hole symmetriy. Template parameters were introduced to enable the same functions to work with objects of different types. Despite the resulting additional complexity, this conveniently enables computations in each of these three formalisms in the same codebase, still using much of the same functionality.

\paragraph{Performance.} Computing dynamical correlation functions is a computationally demanding task, especially for four-point functions which depend on three frequency arguments. Depending on the desired resolution, this requires both excessive memory to store these functions during computations and CPU power to perform computations for each combination of arguments. Concerning the latter, using optimized data structures for efficient readouts of data as well as an efficient but still precise algorithm for integrating over frequencies (the numerical bottleneck) improved matters significantly. Also, using a compiled programming language is basically a must and keeping track of constant variables and member functions helps the compiler optimize the code. 

\paragraph{Scalability.} Apart from the simplest calculations, most diagrammatic calculations would not be feasible without parallelization. This is because practically all calculations in the parquet formalism or mfRG require computations for all possible combinations of external arguments of the correlation functions. As those are independent from each other, it is possible, and advisable, to parallelize the demanding computations of bubbles and loops, see Sec.\,\ref{sec:functionality}, in the external arguments. Using the \code{OpenMP} and \code{MPI} interfaces, this can easily be achieved for parallelization across different threads on the same node and across multiple nodes, respectively. For more details see Sec.\,\ref{sec:parallelization}. As long as the memory requirements are met, the performance of the code scales almost perfectly with the computational resources.\\

At this point we disclose that the present code also has a number of weaknesses, which evolved over the course of development. If the reader intends to set up a new codebase for the purpose discussed here, we recommend considering the following points:

\setcounter{paragraph}{0}
\paragraph{Too many preprocessor macros (``flags'').}\label{par:flags}
The code contains far too many preprocessor macros, used to specify different parameters and settings before compilation, see Sec.\,\ref{sec:parameters}. This not only hampers readability but also increases the risk of errors, as it is never possible to test the full functionality of the code because one would have to compile and test all possible configurations independently. By simple combinatorics, this quickly becomes an overwhelming task. Using preprocessor macros is, however, useful for quick implementations of new functionality, which is why they accumulated over time.

\paragraph{Too many overly complicated structures.}
The code contains several classes that are way more complicated than they need to be, such as the different vertex classes or the data buffer, see Sec.\,\ref{sec:vertex} and Sec.\,\ref{sec:data_structures}. When they were set up, the goal was to keep them as general as possible, such that they could be used for all kinds of models in all kinds of formalisms. For this purpose, templates are used excessively, as well. As a consequence, they are indeed flexible, but they are cumbersome to use in any specific context and their implementations are difficult to grasp. Also, the code takes a long time to compile and link, which is inconvenient for everyday development. Ultimately, as a developer, one has to find the right trade-off between flexibility and simplicity.

\paragraph{Too little use of existing implementations.}
Several textbook algorithms, such as the Gauß-Lobatto routine for frequency integrations, or the Cash-Karp routine for solving ODEs, see Sec.\,\ref{sec:integrations} and Sec.\,\ref{sec:ODE-details}, were implemented by hand. The reason for this was the desire to comprehend and track the inner workings of the algorithms at every point during a calculation. In hindsight, much time and effort could have been saved if existing implementations of these algorithms had been used as ``black boxes''.

\paragraph{Language.}
C++ is a very versatile language, which runs on essentially any computer and can produce very fast code. However, a codebase written in C++ requires a lot of work to write and to maintain. Initially, C++ was chosen for performance reasons. By now, there are, however, established alternative programming languages, that are easier to use, less error-prone, and (almost) as fast, such as Julia\cite{julia}, Rust\cite{matsakis2014rust}, or Mojo\cite{mojo}.

\paragraph{Priorities.}
Driven by the desire to obtain data with maximal resolution and precision, the top priority has always been performance. While this is very typical for codes written by physicists, it is not in line with the typical recommendation in software engineering, which would prioritize correctness and maintainability \emph{over} performance\cite{Stroustrup2023-sj}. While we are confident that the code produces correct results after extensive benchmarks \cite{main_paper}, the code is not written in the simplest way and is not easily readable and maintainable. While we acknowledge that generating results quickly is deemed to be the most important aspect of research nowadays, we advocate for reconsidering the priorities during research software development for future projects. \\

The rest of the paper is structured as follows:
In Sec.\,\ref{sec:model} we briefly introduce the single-impurity Anderson model (AM). In in Sec.\,\ref{sec:methods} we briefly recapitulate the main concepts of diagrammatic many-body theory. In Sec.\,\ref{sec:Keldysh} we comment on the complications that arise by performing computations in the very general Keldysh formalism, which is the main selling point of the present codebase. \\
In the second part of the paper, we give details on the code itself, introducing the main objects in Sec.\,\ref{sec:objects} and explaining the main functionality in Sec.\,\ref{sec:functionality}. We list several options for postprocessing the raw data obtained after a completed calculation in Sec.\,\ref{sec:post-processing} and briefly explain how the data is organized in Sec.\,\ref{sec:io}. Special emphasis is placed on performance-critical aspects of the code in Sec.\,\ref{sec:performance}. We comment on how the code is tested in Sec.\,\ref{sec:testing}. Lastly, we provide an overview over the most important options for parameter choices that can be done in Sec.\,\ref{sec:parameters}, illustrating the versatility of the codebase.\\
In the third main part of the paper, we elaborate on how three different diagrammatic algorithms, perturbation theory, the parquet equations, and the mfRG, are implemented. In particular, we list the different flow schemes that are available in mfRG.
Finally, Sec.\,\ref{sec:conclusion} presents a conclusion.\\

Before the end of this introduction, a disclaimer is in order: This paper does not mention every single class or function in the code, but focuses on the most important aspects and functionalities. Also, while the code enables computations in the KF and the MF at both finite and zero temperature, we focus our specific descriptions mainly on the KF functionality, as this is a unique feature of our codebase.\\

\subsection{Model}\label{sec:model}

We consider the single-impurity Anderson model (AM) in thermal equilibrium, one of the most-studied models in all of condensed matter physics. Its physical behavior is well understood, and numerically exact benchmark data for single-particle correlation functions is available from NRG\cite{Bulla2008} as well as exact analytical results for static quantities at zero temperature from the Bethe ansatz\cite{Zlatic1983, wiegmann_exact_1983}. This makes it an ideal candidate for studies focused on reliable method development.

The AM is a minimal model for localized magnetic impurities in metals introduced by P.W.\,Anderson to explain the physics behind the Kondo effect\cite{anderson_localized_1961}.
It is defined by the Hamiltonian
\begin{align}
H &= 
\sum_{\epsilon\sigma} \epsilon c_{\epsilon\sigma}^\dagger c_{\epsilon\sigma}
+ \sum_\sigma \epsilon_d n_\sigma
+ U n_\uparrow n_\downarrow 
+
\sum_{\epsilon\sigma} (V_\epsilon d_\sigma^\dagger c_{\epsilon\sigma} + \mathrm{H.c.})
,
\label{eq:SIAM_Hamiltonian}
\end{align}
describing a local impurity $d$ level with on-site energy $\epsilon_d$, hybridized with spinful conduction electrons, created by $c_{\epsilon\sigma}^\dagger$, of the metal via a matrix element $V_\epsilon$. Hence, it qualifies as an open quantum system. The electrons in the localized $d$ state, where $n_\sigma \!=\! d_\sigma^\dag d_\sigma$, interact according to the interaction strength $U$, whereas the $c$ electrons of the bath are non-interacting.
The bath electrons are hence formally integrated out, yielding the frequency-dependent retarded hybridization function
$-\mathrm{Im}\, \Delta^R(\nu) = \sum_\epsilon \pi |V_\epsilon|^2 \delta(\nu-\epsilon)$.
We consider a flat hybridization in the wide-band limit, $\Delta^R(\nu) \!=\! -i\Delta$,
so that the bare impurity propagator reads $G_{0}^{R}(\nu) = (\nu - \epsilon_d + i\Delta)^{-1}$.

The code can treat all choices for the on-site energy $\epsilon_d$. For the special choice $\epsilon_d=-U/2$, the model has particle-hole symmetry and is referred to as the symmetric Anderson model (sAM). This setting simplifies the calculations somewhat. For instance, in this case the Hartree-term of the self-energy is constant $\Sigma_\mathrm{H} = U/2$, see also Sec.\,\ref{sec:Hartree}. Also, in the MF, all quantities become real-valued, whereas they are complex-valued otherwise. Hence, the code supplies a parameter flag to make use of these properties, see Sec.\,\ref{sec:parameters}. For general $\epsilon_d\neq-U/2$, we speak of the asymmetric Anderson model (aAM).

Some physical applications require an additional external magnetic field $h$, described by an additional term $h(n_\uparrow - n_\downarrow)$ in the Hamiltonian. At present, the codebase is, however, not applicable in this setting, as this would break SU(2) symmetry, which is heavily used and hard-coded into the codebase, see Sec.\,\ref{sec:symmetries}. A generalization to $h\neq 0$ is possible but would require major effort.

While the present implementation is restricted to the AM, the code in principle can also treat other models: all data structures possess an additional internal index suitable for encoding additional dependencies and quantum numbers of more complicated models, such as a momentum dependence or multiple orbitals. Indeed, first attempts to study the 2D Hubbard model had been started; however, already the simplest KF perturbation theory calculations turned out to be too demanding at the time. The corresponding functionality is therefore not included in this release.

\subsection{Diagrammatic many-body theory}\label{sec:methods}

The basic objects of interest in all our calculations are one- and two-particle correlation functions. Their non-trivial contributions due to interaction effects are contained in the self-energy $\Sigma$ and the four-point vertex $\Gamma$,
\begin{align}
    \Sigma =
    \tikzm{selfenergy}{
        \selfenergywithlegs{$\Sigma$}{0}{0}{1}
    }\, , \qquad
    \Gamma = 
    \tikzm{vertex}{
        \fullvertexwithlegs{$\Gamma$}{0}{0}{1}
    }
    \, .
\end{align}
The self-energy is used together with the bare propagator $G_0$ to express the one-particle propagator $G$ via the Dyson equation,
\begin{align}
    G \,
    = \,
    \begin{gathered}
    \tikzm{Keldysh_formalism-Dyson_G}{
        \draw[lineWithArrowCenter] (1,0) -- (0,0);
    }
    \end{gathered}
    \, = \,
    \begin{gathered}
    \vspace{-2.5ex}
    \tikzm{Keldysh_formalism-Dyson_G0}{
        \draw[lineBareWithArrowCenter] (1,0) -- (0,0);
        \node at (0.5,-0.4) {$G_0$};
    }
    \end{gathered}
    \, + \,
    \begin{gathered}
    \vspace{-1.5ex}
    \tikzm{Keldysh_formalism-Dyson_G0SigmaG}{
        \draw[lineBareWithArrowCenter] (1,0) -- (0,0);
        \selfenergy{$\Sigma$}{1.3}{0}{1}
        \draw[lineWithArrowCenter] (2.6,0) -- (1.6,0);
        \node at (0.5,-0.4) {$G_0$};
        \node at (2.1,-0.4) {$G$};
    }
    \end{gathered}
    \, , \label{eq:Dyson}
\end{align}
which is formally solved by $G = 1/(G_0^{-1} - \Sigma)$. The vertex is the connected and amputated part of the two-particle correlation function $G^{(4)}$,
\begin{align}
    G^{(4)}
    \, =\  
    \tikzm{Keldysh_formalism-G4_dcon1}{
        \draw[lineWithArrowCenter] (0.8,-0.4) -- (0,-0.4);
        \draw[lineWithArrowCenter] (0,0.4) -- (0.8,0.4);
    }
    \ \ -\ \
    \tikzm{Keldysh_formalism-G4_dcon2}{
        \draw[lineWithArrowCenter] (0.8,-0.4) -- (0.8,0.4);
        \draw[lineWithArrowCenter] (0,0.4) -- (0,-0.4);
    }
    \ \ + \ \
    \tikzm{Keldysh_formalism-G4_con}{
        \fullvertex{$\Gamma$}{0}{0}{4./3.};
        \draw[lineWithArrowCenter] (-0.4,-0.4) -- (-1.2,-0.4);
        \draw[lineWithArrowCenter] (1.2,-0.4) -- (0.4,-0.4);
        \draw[lineWithArrowCenter] (0.4,0.4) -- (1.2,0.4);
        \draw[lineWithArrowCenter] (-1.2,0.4) -- (-0.4,0.4);
    }
    \ \ ,
\end{align}
from which physical susceptibilities can be obtained by contracting pairs of external legs (see. App.\,C of \onlinecite{main_paper} for details). The first-order contribution to the vertex is given by the fully antisymmetric, local, and instantaneous bare vertex, represented as a single dot, 
\begin{align}
    \Gamma_0 \ = \
    \tikzm{Keldysh_formalism-Gamma0}{
        \barevertexwithlegs{0}{0}
    }
     \ \sim \ U >0\, ,
\end{align}
in standard Hugenholtz notation. Using the bare vertex and the bare propagator $G_0$, diagrammatic perturbation series for both the self-energy and the vertex can be derived, which will be the subject of Sec.\,\ref{sec:perturbation_theory}. A perturbation series up to finite order in $\Gamma_0$ is, however, only appropriate for weak coupling strengths. In order to reach larger couplings, an infinite number of diagrams has to be summed. This is the purpose of two related formalisms, the parquet formalism and the multiloop functional renormalization group, to be discussed in Sec.\,\ref{sec:parquet} and Sec.\,\ref{sec:mfRG}, respectively. Both formalisms employ the parquet decomposition to organize all diagrammatic contributions to $\Gamma$ into one of four distinct categories: Two-particle reducible diagrams in one of the three two-particle channels $a$, $p$ and $t$, included in the three two-particle reducible vertices $\gamma_{r\in\{a,p,t\}}$ or two-particle irreducible diagrams, included in the fully irreducible vertex $R$,
\begin{subequations}
\begin{align}
    \Gamma
    &= 
    \tikzm{R}{
        \fullvertexwithlegs{$R$}{0}{0}{1}
    }
    + 
    \tikzm{gamma_a}{
        \fullvertexwithlegs{$\gamma_a$}{0}{0}{1}
    }
    +
    \tikzm{gamma_p}{
        \fullvertexwithlegs{$\gamma_p$}{0}{0}{1}
    }
    +
    \tikzm{gamma_t}{
        \fullvertexwithlegs{$\gamma_t$}{0}{0}{1}
    } \\
    &= 
    \tikzm{Keldysh_formalism-Gamma0}{
        \barevertexwithlegs{0}{0}
    }
    + 
    \tikzm{mfRG-parquet-SDE0}{
            \abubblebarebarebare{0}{0}{0.85};
		\barevertexlefthalf{0}{0};
		\barevertexrighthalf{1}{0};
		}
    + \tfrac{1}{2}
    \tikzm{mfRG-parquet-SDE0}{
            \pbubblebarebarebare{0}{0}{0.85};
		\barevertex{0}{0};
		\barevertex{1}{0};
		\arrowslefthalfp{0}{0}{0.8};
		\arrowsrighthalfp{1}{0}{0.8};
		\node at (0,0.6){};
		}
    -
    \begin{gathered}
        \tikzm{mfRG-parquet-SDE0}{
            \tbubblebarebarebare{0}{0}{0.8};
		\barevertexupperhalf{0}{-1};
		\barevertexlowerhalf{0}{0};
		}
    \end{gathered}
    + O[(\Gamma_0)^3]\, .
\end{align}
\end{subequations}
Any specific diagram is said to be two-particle reducible if it can be disconnected by splitting a propagator pair. Otherwise, it is said to be two-particle irreducible.
The parquet decomposition is exact, as it in essence just provides a classification of all diagrams that contribute to $\Gamma$. However, neither the parquet formalism nor the mfRG provide equations for $R$. In practice, some approximation is required. The simplest one is the parquet approximation (PA),
\begin{align}
    R = \Gamma_0 + O[(\Gamma_0)^4] \approx \Gamma_0,
\end{align}
which approximates the fully irreducible vertex $R$ by the bare vertex $\Gamma_0$. As it introduces an error in the fourth order in perturbation theory, it fails for large coupling strengths and is hence applicable only up to intermediate couplings.
The PA was applied throughout in Ref.\,\onlinecite{main_paper} and is the only one so far implemented in the codebase (see Sec.\,\ref{sec:vertex} for a comment on other possibilities.)

\subsection{Keldysh formalism}\label{sec:Keldysh}

The following section assumes familiarity with the KF and describes challenges arising for computations with the KF rather than the more widespread MF. For a more extensive discussion of the KF, see Refs.\,\onlinecite{Kamenev_2023, walter_keldyshfrg_2022}.

The KF\cite{keldysh_diagram_1965, schwinger_brownian_1961, kadanoff_quantum_1962} works both out of equilibrium and in thermal equilibrium at arbitrary temperature, in a real-frequency description. This is an advantage over the more popular MF, which works in imaginary (``Matsubara'') frequencies, requiring analytical continuation; a mathematically ill-defined problem if one works with a finite amount of imperfect numerical data. Still, the KF is seldomly used, because practical calculations are more complicated for two main reasons:

In the KF, all operators acquire an additional contour index, which specifies whether they sit on the forward or backward branch of the Keldysh double-time contour. It follows that the four-point vertex, for example, has $2^4=16$ different components. While some of these components can be eliminated by causality or related to other components by fluctuation-dissipation relations in thermal equilibrium or symmetries, this additional index structure complicates the implementation and the numerics. 

In thermal equilibrium, energy conservation can be leveraged by Fourier-transforming all correlation functions to frequency space. In contrast to the MF at finite temperatures, this dependence is continuous. Hence, contractions over frequency arguments require numerically more expensive integrations instead of summations. The integrations become more costly at lower temperatures, as the frequency dependence of the correlation functions becomes more sharply peaked. The four-point functions, which depend on three continuous frrequency arguments, are the numerical bottleneck for which arbitrarily high resolutions are out of reach due to both computation and memory demands. Discretizing the frequency dependence in a clever way and using adaptive integration routines is therefore key, as discussed in Sec.\,\ref{sec:frequency_grid} and Sec.\,\ref{sec:integrations}.

Lastly, the KF also allows for computations out of thermal equilibrium. However, the present discussion is restricted to thermal equilibrium. Extending the code out of equilibrium is possible with moderate effort.

\section{The Code} 

In part II of the paper, we describe the main building blocks of the code -- the classes representing correlation functions and other functions for combining them in diagrammatic computations. Furthermore, we describe post-processing schemes and emphasize aspects important for performance. More information on technical details of individual code pieces can be found in the documentation attached to the source code, see the code availability statement at the end of this paper.

\subsection{Prerequisites}

The code itself is written in \code{C++17} \cite{cpp17standard} and is built using \code{CMake} \cite{cmake}, demanding at least version $3.10$. It requires the \code{GSL} \cite{galassi_gnu_2009}, \code{boost} \cite{boost} and \code{Eigen3} \cite{eigenweb} libraries as well as the \code{HDF5} \cite{hdf5} library for input and output. For parallelization, the \code{OpenMP} \cite{openmp} and \code{MPI} \cite{MPI} interfaces are used. Notably, we do not supply precompiled executables, that could be run directly, for several reasons: First, the code makes heavy use of preprocessor flags that must be set \emph{before} compilation and that are in part used to specify the concrete problem at hand, see Sec.\,\ref{sec:parameters}. Second, special compilers for the particular architecture at hand might be available, which could optimize the code during compilation and linking, improving the performance. The user should hence adapt the file \code{CMakeLists.txt} accordingly, such that the required libraries are included and linked properly and all compiler settings are as desired.

The technical documentation supplied with the code is generated automatically using the tools Doxygen\cite{doxygen}, Sphinx\cite{sphinx}, Breathe\cite{breathe}, and CMake.

\subsection{Basic structure}

\begin{figure}
    \begin{tikzpicture}
        \node at (-0.7,-0.5) {\boxed{\code{FrequencyGrid}}};
        \node at (-1.5,-1.4) {\code{State}};
        \node at (-1.2,-2) {\boxed{\code{Vertex}\ \Gamma}};
        \node at (1.2,-2) {\boxed{\code{SelfEnergy}\ \Sigma}};
        \draw[dashed] (-2.2, -2.5) rectangle (2.6, -1);
        \node at (4.5,-2) {\boxed{\code{Propagator}\ G, \dot{G}, S}};
        \node at (4,-3) {\boxed{\code{Bubble}\ \Pi, \dot{\Pi}}};
        \node at (-0.5,-3) {\boxed{\code{bubble\_function}}};
        \node at (-1.5,-3.8) {\boxed{\code{loop}}};
        \draw[->, thick] (-0.8, -0.75) -- (-0.8, -1.75);
        \draw[->, thick] (0.3, -0.75) -- (0.3, -1.7);
        \draw[->, thick] (2.4, -2) -- (2.9, -2);
        \draw[->, thick] (4, -2.3) -- (4, -2.7);
        \draw[->, thick] (2.95, -3) -- (1, -3);
        \draw[->, thick] (-1.2, -2.75) -- (-1.2, -2.25);
        \draw[->, thick] (-2, -2) -- (-2.5, -2) -- (-2.5, -3.8) -- (-2, -3.8);
        \draw[->, thick] (-2.5, -3) -- (-2, -3);
        \draw[->, thick] (5.5, -2.3) -- (5.5, -4.3) -- (-1.5, -4.3) -- (-1.5, -4.05);
        \draw[thick] (-1, -3.8) -- (1.8, -3.8) -- (1.8, -3.1);
        \draw[->, thick] (1.8, -2.9) -- (1.8, -2.3);
    \end{tikzpicture}
    \caption{Schematic depiction of the main parts of the codebase.}
    \label{fig:main}
\end{figure}
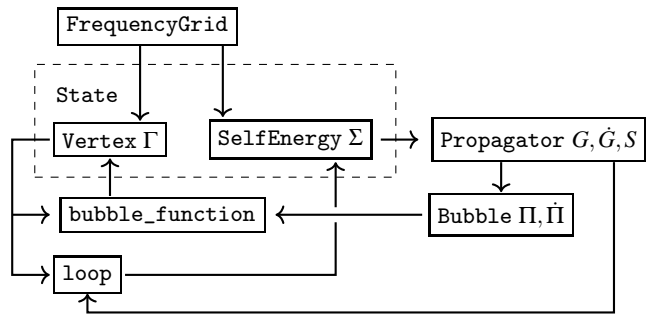

The structure of the main part of the codebase is depicted in Fig.~\ref{fig:main}.
The main objects of interest are the \code{SelfEnergy} $\Sigma$ and the four-point \code{Vertex} $\Gamma$. Separate classes have been implemented for both, discussed in detail below. Both classes use instances of the class that defines suitably chosen \code{FrequencyGrid}s, to be discussed in Sec.\,\ref{sec:frequency_grid}, for discretizing the continuous frequency dependence. A self-energy and a vertex always come together in any practical calculation, representing data for a step of an mfRG flow or an iteration of the parquet solver. The self-energy and vertex classes are hence combined in a \code{State} class $\Psi = (\Sigma, \Gamma)$. The algorithms discussed in Sec.\,\ref{sec:algorithms} require computing bubble- and loop-type diagrams, the main functionality of the codebase. As detailed in Sec.\,\ref{sec:bubble_function} below, the \code{bubble\_function} contracts two input vertices with a pair of propagators in one of the three two-particle channels, to yield a new four-point vertex, which is stored as an instance of the \code{Vertex} class. For example, contracting two vertices $\Gamma_1$ and $\Gamma_2$ in the $a$ channel is denoted as 
\begin{align}
    B_a(\Gamma_1, \Gamma_2) = \Gamma_1 \circ \Pi_a \circ \Gamma_2 = 
    \tikzm{Keldysh_vertex_parametrization-channeldec-Ba}{
		\arrowslefthalffull{0}{0}{1}
		\fullvertex{$\Gamma_1$}{0}{0}{1}
		\abubblefullfull{0.3}{0}{1}{1}{1}
		\fullvertex{$\Gamma_2$}{1.8}{0}{1}
		\arrowsrighthalffull{1.8}{0}{1}
	}
    \, ,
\end{align}
see also App.\,C in Ref.\,\onlinecite{main_paper} for a fully parametrized version.
The required propagator pair $\Pi$ belongs to a separate \code{Bubble} class, ensuring the correct combination of propagators and their parametrization. The propagators themselves are defined in the \code{Propagator} class, which essentially implements the Dyson equation, Eq.\,\eqref{eq:Dyson}, combining $G_0$ and $\Sigma$. The former contains all the system parameters, including the regulator in mfRG,
the latter encodes the interaction effects. Both the \code{Propagator} and \code{Bubble} classes can handle differentiated objects, arising in mfRG, see Sec.\,\ref{sec:mfRG}.
Lastly, the \code{loop} function is used to contract two external legs of a four-point \code{Vertex} with a \code{Propagator}, yielding an instance of the \code{SelfEnergy} class, for example
\begin{align}
    L(\Gamma, G) = 
    \tikzm{mfRG-1l-dSigma}{
			\fullvertex{$\Gamma$}{0}{0}{1}
			\arrowslowerhalffull{0}{0}{1}
			\loopfullvertex{lineWithArrowCenter}{0}{0}{1}
		}
  \, .
\end{align}
These types of diagrams are required, e.g., for the mfRG flow equation of the self-energy, or for the evaluation of the SDE after a previous bubble diagram computation.

\subsection{Correlation function classes}\label{sec:objects}

In the following, we discuss the main building blocks of the code in more detail.
We begin by outlining the self-energy and vertex classes. In addition, there are two helper classes: the first represents propagators, combining the bare propagator and the self-energy, the second combines a pair of propagators as needed for bubble-type diagrams.

\subsubsection{The \code{Vertex} classes}\label{sec:vertex}

In total, the code contains the four classes \code{irreducible}, \code{rvert}, \code{fullvert}, and \code{GeneralVertex} to store different types of four-point vertices. 

The \code{irreducible} class contains the two-particle irreducible part of the vertex, $R$. In the PA, its 16 Keldysh components are just constants. It can easily be extended to hold nontrivial input data, for example in the context of diagrammatic extensions\cite{Rohringer2018} of dynamical mean-field theory\cite{Georges1996} such as D$\Gamma$A\cite{Toschi2007} or DMF$^2$RG\cite{taranto_infinite_2014}.

The \code{rvert} class stores the two-particle reducible vertices $\gamma_{r\in \{a,p,t\}}$. Each of them is split up into their asymptotic classes\cite{wentzell_high-frequency_2020} $K_1$, $K_2$ and $K_3$, where the $K_{2'}$ class is inferred from $K_2$ by crossing symmetry. Being one-, two- and three-dimensional objects, respectively, each of those naturally has its own frequency grid. The \code{rvert} supplies several methods to store and read out data, either directly or interpolated. Conveniently, it can return all vertex parts where external legs either do or do not meet at the same bare vertex on the left or on the right-hand side, by suitably combining the $K_1$, $K_{2^{(')}}$ or the $K_{2^{(')}}$ and $K_3$ classes, respectively. This turned out to be very handy for keeping track of contributions for the different asymptotic classes during calculations. In addition, the \code{rvert} class can track and, if desired, enforce symmetries, in the Keldysh-, spin- and frequency domains, see Sec.\,\ref{sec:symmetries} for details. For debugging purposes, functionality not using symmetries is provided as well.

The \code{fullvertex} class combines one instance of the \code{irreducible} class and three instances of the \code{rvert} class, one for each two-particle channel $a, p, t$. It can then return the value of the full vertex, which is the sum of the four contributions, for a given Keldysh and spin component, interpolated at a given combination of frequencies. As each individual \code{rvert} instance, it can collect all those parts of the vertex where the external legs either do or do not meet at the same bare vertex on the left or on the right-hand side and includes functionality to exploit various symmetries. In addition, it can compute the $p$-norm of each asymptotic contribution, which is useful for debugging purposes and convergence criteria, e.g.~in parquet computations.

While instances of the \code{fullvertex} class hold the data of the symmetry-reduced sector of a full vertex, certain diagrammatic equations involve subsets of vertex diagrams. One example is the $r$-channel-irreducible vertex used in the Bethe-Salpeter equations outlined in Sec.\,\ref{sec:parquet}. Such diagrams do not necessarily obey all the symmetries of a full vertex, so they must be treated differently. These asymmetric cases are therefore encoded in the \code{GeneralVertex} class. It uses multiple instances of \code{fullvertex} which together cover the symmetry-reduced sector of the asymmetric vertex data. Let us comment here that, while this approach is feasible, it turned out to be inconvenient in practice, as one always has to make sure that all sectors are covered, i.e. that all required \code{fullvertex} instances are provided. This is a source of logical errors that can sometimes be hard to find. In retrospect, it would have been better to pay the increased cost in memory to store all vertex contributions in the same object, making the code easier to read and to work with.

All vertex classes allow adding or subtracting two instances of the respective classes, or multiplying a number with a vertex instance.

Splitting up the vertex functionality into so many different classes was made in the beginning of developing the code to provide enough flexibility, in particular regarding symmetries and a possible non-trivial input for the irreducible vertex. In hindsight, it turned out that for the computations done in Ref.\,\onlinecite{main_paper}, this structure would not have been required in this generality.

\subsubsection{The \code{SelfEnergy} class}

The \code{SelfEnergy} class comes with a \code{dataBuffer} that stores the discrete values of the retarded and Keldysh components of the self-energy on a given frequency grid, see Sec.\,\ref{sec:frequency_grid} and Sec.\,\ref{sec:data_structures}. When instantiating an object of the \code{SelfEnergy} type, a given frequency grid can either be supplied, or a suitable one is generated automatically based on the value of the regulator $\Lambda$. In addition, the asymptotic value of the retarded component of the self-energy has to be set. Most of the time, this should be the Hartree value $\Sigma_\mathrm{H}$, as the \code{SelfEnergy} inside the code is supposed to be used only for the dynamical, i.e.~frequency-dependent, contributions of the self-energy, which excludes the constant Hartree value. For the sAM, the Hartree value is constant, $\Sigma_\mathrm{H}=U/2$, in the asymmetric case it has to be computed self-consistently beforehand. This can be done inside the code using the \code{HartreeSolver} class, see Sec\,\ref{sec:Hartree}.

The \code{SelfEnergy} class provides a host of methods used throughout the code. Most importantly, it can return the value of the self-energy either directly at a given input on the frequency grid (fast) or return an interpolated value at a given continuous frequency (not so fast). It can also set the value of $\Sigma$ to a given input. In addition, one can compute the $p$-norm of $\Sigma$, and the relative deviation to a different \code{SelfEnergy} instance, using the maximum norm . This is used to check convergence in parquet computations detailed in Sec.\,\ref{sec:parquet}.

Lastly, multiple operators are defined for the \code{SelfEnergy} class, which are used to add or subtract two \code{SelfEnergy} instances or to multiply some number with a \code{SelfEnergy} instance.

\subsubsection{The \code{State} class}\label{sec:state}

Instances of the \code{State} class are the high-level objects that are mainly used by the high-level algorithms discussed in Sec.\,\ref{sec:algorithms}. The \code{State} class combines a \code{GeneralVertex} and a \code{SelfEnergy}, which together contain all non-trivial information that one might wish to compute. In that sense, they suffice to completely specify the ``state'' of the calculations. For the purpose of fRG calculations, the \code{State} class also holds the value of the flow parameter $\Lambda$.

As the vertex classes and the \code{SelfEnergy} class, the \code{State} class also comes with operators that can be used to add and subtract states from one another and to multiply a number with a state. These operators under the hood just invoke the corresponding operators previously defined for the vertex and self-energy. Hence, all high-level algorithms can manipulate instances of the \code{State} class directly, e.g.~combining several iterations of the parquet solver in a mixing scheme.

\subsubsection{The \code{Propagator} class}

The \code{Propagator} class is special in the sense that it stores almost no data itself. Instead, it references instances of the \code{SelfEnergy} class and combines the analytical form of the bare propagator $G_0$ with the self-energy via the Dyson equation, $G = 1 / [(G_0)^{-1} - \Sigma]$. To that end, it can return the value of a given propagator at some argument, interpolated on the frequency grid of the referenced self-energy. This can be done either directly for a given Keldysh component at some continuous frequency or vectorized over all Keldysh components. As $G_0$ depends on the formalism used and in  mfRG on the choice of the regulator, separate methods for a variety of choices are provided. In addition, one can specify whether the full propagator $G$ shall be computed, or the single-scale propagator $S$, the differentiated propagator including the Katanin extension \cite{katanin_fulfillment_2004}, or just the Katanin extension by itself, see Sec.\,\ref{sec:mfRG}. Note that the Katanin extension requires the self-energy differentiated with respect to the flow parameter $\Lambda$, hence the propagator class references \textit{two} \code{SelfEnergy} instances, one non-differentiated, and one differentiated.

\subsubsection{The \code{Bubble} class}

Lastly, the \code{Bubble} class combines two propagators to yield a bubble in one of the three two-particle channels $a$, $p$ and $t$, according to Eqs.\,(C1a-c) in Ref.\, \onlinecite{main_paper}. For evaluating differentiated bubbles in mfRG, one of the propagators can be chosen to be the single-scale propagator $S$ or the fully differentiated one $\dot{G}$. In that case, the bubble already takes care of the product rule, giving (symbolically) $\dot{\Pi}^S = G S + S  G$ or $\dot{\Pi} = G\dot{G} + \dot{G} G$. Otherwise, it just yields $\Pi = G G$. 
The \code{Bubble} class provides functions for obtaining the value of a bubble in a given channel at specified bosonic and fermionic frequencies, either for one specific Keldysh component directly or vectorized over the Keldysh structure. This class simplifies bubble computations using the \code{bubble\_function}, see Sec.\,\ref{sec:bubble_function}.

\subsection{Main functions for diagrammatic computations}\label{sec:functionality}

Computing bubbles and loops involves contractions over quantum numbers and Keldysh indices, including integrations over frequencies for all possible combinations of external arguments, and is by far the most costly part for the numerics. A clean and efficient implementation of this functionality is therefore paramount and should be of the highest priority when setting up a new code. In the following, we provide technical details on this most important part of the code.

\subsubsection{The \code{bubble\_function}}\label{sec:bubble_function}

The \code{bubble\_function} implements equations (C2a-c) from Ref.\,\onlinecite{main_paper}. It takes references to three vertices as arguments, one to store the result of the computation and two others to be connected by a \code{Bubble} object. This \code{Bubble} object can either be supplied as well or is initialized by an overload of the \code{bubble\_function}, which in addition requires the two propagators that shall be used for the \code{Bubble}. The main work is then done by an instance of the class \code{BubbleFunctionCalculator}, which performs the bubble contractions for each diagrammatic class separately. This is done for every possible combination of external arguments, i.e.~Keldysh indices and frequencies. At this point, the calculations are parallelized as outlined in Sec.\,\ref{sec:parallelization}. For each set of arguments, an \code{Integrand} object is instantiated, which puts together the two vertices and the bubble and performs the contraction over Keldysh indices, if the flag \code{SWITCH\_SUM\_N\_INTEGRAL} is set to 1. The \code{Integrand} class provides an operator that reads out the integrand at a given frequency. It is called by the integrator, invoked subsequently and described in detail in Sec.\ref{sec:integrations}. The results of all the frequency integrations are finally collected and \emph{added} to the vertex object that was given as the first argument to the \code{bubble\_function}. The choice not to output a completely new vertex but instead add the result to an existing vertex has historical reasons to save memory. This increased the risk of logical errors during high-level algorithm implementations, though, and in hindsight, the \code{bubble\_function} should better have been designed to output a completely new vertex object.

\subsubsection{The \code{loop} function}\label{sec:loop}

The \code{loop} function implements equation (C3) from Ref.\,\onlinecite{main_paper} and is structured similarly to the \code{bubble\_function}. It takes a reference to a self-energy for storing the result as well as references to a vertex and a propagator as arguments of the loop. For each external fermionic frequency, in which the computation is parallelized again, it invokes the integrator to perform a frequency integration using the \code{IntegrandSE} class. For the aAM, the asymptotic value of the just computed self-energy is extracted from the Hartree- and the $K_{1,t}$ and $K_{2',t}$ terms after the calculation. For the sAM, the asymptotic value of the self-energy is a known constant.

\subsection{Postprocessing}\label{sec:post-processing}

The code provides a host of postprocessing functions. These are not required for the actual calculations themselves but are useful to extract additional information from their results, either as consistency checks or to infer derived quantities for later analysis.

\subsubsection{Causality check for the self-energy}

By causality, the imaginary part of the retarded component of the self-energy is strictly non-positive\cite{Kugler2022}, $\mathrm{Im}\,\Sigma^R(\nu) \leq 0$ for all frequencies $\nu\in \mathbb{R}$. A violation of this condition not only constitutes an unphysical result but often leads to numerical instabilities. The code therefore provides the function \code{check\_SE\_causality} that checks this condition for a supplied instance of \code{SelfEnergy}. Typically, this function is invoked after each ODE step during an mfRG calculation or after each iteration of the parquet solver.

\subsubsection{Fluctuation dissipation relations}

In thermal equilibrium at temperature $T$, one has a fluctuation-dissipation relation (FDR)\cite{jakobs_functional_2010, Kamenev_2023} between the retarded and the Keldysh components of the propagator, $G^K(\nu) = 2i\tanh(\tfrac{\nu}{2T})\, \mathrm{Im}\, G^R(\nu)$, and the self-energy, $\Sigma^K(\nu) = 2i \tanh(\tfrac{\nu}{2T})\, \mathrm{Im}\Sigma^R(\nu)$.
This relation can be used to infer the Keldysh components of the self-energy from the retarded component or vice versa, hence it would in principle suffice to compute only one of the components. However, in the vectorized form of the code, both components of the self-energy are computed anyway. The FDR can hence be used as an internal consistency check, provided by the function \code{check\_FDTs\_selfenergy}. It computes $\Sigma^K$ from $\Sigma^R$ via the FDR and compares it to the independently computed Keldysh-component of the self-energy by computing the $2$-norm of the difference.

As an additional consistency check, the fulfillment of fluctuation-dissipation relations for the $K_1$ classes, reading
\begin{align}
    \mathrm{Im}\,K_1^R(\omega) &=
        -\frac{i}{2} \tanh\left(\tfrac{\omega}{2T}\right) K_1^K(\omega)
\end{align}
can be examined. One may also want to check generalized FDRs for three-point and four-point contributions of the vertex \cite{wang_generalized_2002}.

\subsubsection{Kramers-Kronig relation}

For functions $f(\omega)$ that are analytic in the upper half plane, like retarded single-particle correlation functions, the Kramers-Kronig transform relates the real and imaginary parts via
\begin{align}
    \mathrm{Re}\, f(\omega) = \frac{1}{\pi} \mathcal{P} \int_{-\infty}^\infty \mathrm{d}\omega'\, \frac{\mathrm{Im}\, f(\omega')}{\omega' - \omega},
\end{align}
where $\mathcal{P}$ denotes the Cauchy principal value. Inside the code, the function \code{check\_Kramers\_Kronig} can be used to test how well this generic analytic property is fulfilled.

\subsubsection{Sum rule for the spectral function}

The fermionic spectral function $A(\nu) = -\mathrm{Im}\, G^R(\nu) / \pi$ must obey the sum rule 
\begin{align}
    \int_{-\infty}^\infty \mathrm{d}\nu\, A(\nu) = 1.
\end{align}
The function \code{sum\_rule\_spectrum} implements this integral as a consistency check.

\subsubsection{Susceptibilities}

Susceptibilities, which are of significant physical relevance, are derived from the vertex by contracting pairs of external legs. Diagrammatically, the formula for the $a$-channel susceptibility reads
\begin{align}
    \chi_a &= 
    \tikzm{mfRG-parquet-SDE0}{
		\abubblebarebare{0}{0}{1};
		}
    \,
    +
    \,
    \tikzm{mfRG-parquet-SDE0}{
	\abubblebarefull{0}{0}{1}{1}
	\fullvertex{$\Gamma$}{1.5}{0}{1}
	\abubblefullbare{1.8}{0}{1}{1}
		}
    \, ,
\end{align}
and similarly for the susceptibilities in the $p$ and $t$ channel. 
The fully parametrized equations are provided in Eqs.~(C7) of Ref.\,\onlinecite{main_paper}. Linear combinations of these diagrammatic susceptibilities yield the physical susceptibilities, see Eqs. (C8) of Ref.\,\onlinecite{main_paper}. The code computes susceptibilites using the function \code{compute\_postprocessed\_susceptibilities}, which can be invoked after a completed calculation using the name of the file that stores the results. It iterates through all layers that correspond to ODE steps or parquet iterations (see Sec.\,\ref{sec:io}), evaluates Eqs.\,(C7) using the vertex and self-energy for each and stores the results as a new dataset into the same file.

It was found in Ref.\,\onlinecite{wentzell_high-frequency_2020} that for converged parquet computations, susceptibilities can more easily be extracted directly from the $K_1$ class. As discussed in Refs.\,\onlinecite{main_paper, PhysRevResearch.2.033372}, one can also choose to compute susceptibilities that way during fRG computations, even though the two schemes are inequivalent if multiloop convergence is not reached. The two different schemes of computing susceptibilities can then be used to gauge the quality of the truncation.

\subsubsection{Vertex slices}

Lastly, the function \code{save\_slices\_through\_fullvertex} can be used to read out two-dimensional ``slices'' of the full vertex. It takes the filename corresponding to the results of a finished calculation as an argument, iterates through all layers and saves a two-dimensional cut of all Keldysh components of the full vertex in the $t$-channel parametrization for zero transfer frequency $(\omega_t=0, \nu_t, \nu'_t)$ for a given spin component. While this function does not perform any non-trivial calculations, it is useful for visualization purposes. If desired, the function can be straightforwardly adapted to store vertex slices at finite transfer frequencies, enabling full scans through the three-dimensional structure of the four-point vertex.

\subsection{I/O}\label{sec:io}

We use the HDF5 file format \cite{hdf5} for input and output purposes throughout. To organize the data for output, the contents of a state are split into different datasets that correspond, e.g., to all the asymptotic classes of the vertex in each channel, the self-energy, the frequency grids and the most important parameters of the calculation. The output file is then organized on a high level in terms of ``$\Lambda$ layers'', the idea being that each layer enables access to a different state stored in the same file. Thereby, a single file contains, e.g., the results of a full mfRG flow, where each ``$\Lambda$ layer'' corresponds to a different value of the regulator. Alternatively, this structure can be used to store the results of all iterations needed for solving the parquet equations. Of course, one can equally well use just a single layer to store the end result of a computation, such as a converged solution of the parquet equations or the result of a PT2 computation.

The function \code{write\_state\_to\_hdf} creates a new file with a fixed number of layers and saves an initial state into the first layer. Additional states generated during subsequent computations can be added to the same file (but into a different layer to be specified) using the function \code{add\_state\_to\_hdf}. In effect, these functions are wrappers of a host of additional functions, which are able to store various data structures, such as scalars, vectors, or even \code{Eigen}-matrices to an HDF file.

When using parallelization, as detailed in Sec.\,\ref{sec:parallelization}, one has to ensure that only one single process writes data into the output file. Collisions, where multiple processes simultaneously try to write to the same location in memory, will cause the program to crash.

It is possible to read data from an existing HDF file to generate a new state for subsequent computations. For this purpose the function \code{read\_state\_from\_hdf} reads a state from a specified layer of a provided HDF file. One can thus do checkpointing: If all steps of an mfRG flow or all iterations of the parquet solver are stored separately, a computation that was interrupted can be continued from the last step stored. This design feature is useful for large computations that have to be split over several separate jobs or in the case of a hardware error causing a job to crash. Setting up checkpointing functionality is therefore strongly recommended.

\subsection{Performance}\label{sec:performance}

In the following, we discuss parts of the code of special importance for performance. Of course, there is always a trade-off between accuracy and performance, as, e.g., an arbitrary high frequency resolution quickly becomes prohibitive. Nevertheless, efficient implementations are necessary for challenging computations. For the precision-focused calculations for which this codebase was developed, these parts are therefore of utmost importance.

\subsubsection{\code{MPI}+\code{OpenMP} parallelization}\label{sec:parallelization}

As mentioned in the beginning, mfRG and parquet computations can be heavily parallelized, since the correlation functions are (repeatedly) evaluated independently for every possible combination of external arguments. Parallelization is especially advisable for computing bubbles of two four-point vertices as outlined in Sec.\,\ref{sec:bubble_function}. We use the \code{OpenMP} interface for parallelization across multiple threads on a single node, and the \code{MPI} interface for parallelization across multiple nodes. While \code{OpenMP} parallelization works with shared memory, meaning that all threads have access to the same data on the node that they are running on, one has to be careful with \code{MPI} parallelization working on distributed memory. Processes that run on different nodes to compute, say, a four-point vertex for different sets of external arguments, cannot write their results into the same instance of a four-point vertex. Hence, we introduce additional buffers distributed across the nodes. After the computation of, say, a four-point vertex is finished, these buffers are collected and their contents are put together to yield the full result. While this scheme is initially somewhat cumbersome to set up, it pays off tremendously, as the code's performance scales well with the computational resources, including multiple nodes. This is because, first, computations for different external arguments are independent from each other, so there is minimal communication between the nodes. Second, the number of external arguments required for precision-focused calculations is large, so that individual threads have little downtime waiting for other threads to finish. For example, the most expensive calculations of Ref.\,\onlinecite{main_paper} involved 125 points along each of the three frequency axes and were parallelized across 32 nodes running 32 threads each. Provided enough CPU power, the resolution could in principle be increased further, but is ultimately limited by memory.

\subsubsection{Vectorization}

As outlined in Sec.\,\ref{sec:Keldysh}, KF calculations require computing $2^n$ Keldysh components of $n$-point functions. These components can be arranged into a matrix, yielding, e.g., a $4\times 4$ matrix for the four-point vertex. This structure can be exploited for summing over Keldysh indices by using vectorization and the data structures of the \code{Eigen} library\cite{eigenweb}, significantly improving performance. 
This works because all Keldysh components are stored in contiguous sections of memory. Of course, the other parts of the code have to be able to use these data structures properly, which is why all functions that enable, e.g., access to the correlation functions (see Sec.\,\ref{sec:objects}) have two versions: One that can handle matrix-valued data when vectorization is used, and another used otherwise.

When using vectorization, all Keldysh components have to be stored explicitly. As a consequence, identities that relate different Keldysh components, such as certain symmetries or FDRs cannot be used to reduce the numerical effort. Although maximal exploitation of symmetries initially was one of our main objectives, we later found that vectorization over Keldysh components is preferential despite the larger memory costs.

In the finite-$T$ MF, we use vectorization to represent the Matsubara frequency dependence of all correlation functions. This leads to massive speedups when performing Matsubara sums as matrix-multiplications.

\subsubsection{Symmetries}\label{sec:symmetries}

Many symmetries for reducing the number of data points to be computed directly can still be used together with vectorization over Keldysh indices. These include crossing symmetry of the vertex, which relates a vertex to itself with one pair of external fermionic legs exchanged, complex conjugation of the vertex, SU(2) symmetry in the absence of a magnetic field (which in combination with crossing symmetry reduces the number of independent spin components to 1) and frequency symmetries in the presence of particle-hole symmetry. For explicit details on these symmetries, see App.\,A in Ref.\,\onlinecite{main_paper}.

Since frequency integrations are the most costly part of the computations, symmetry operations are not used for evaluating integrands on the fly. Instead, they are used to reduce the number of vertex components to be computed. Since the vectorized version of the code performs sums over Keldysh indices by matrix multiplication, the result of the integration contains all Keldysh components. Hence, we use the symmetry relations to reduce the other arguments, i.e.\, spin and frequency. The information about the symmetry-reduced components is encoded in symmetry tables. These contain entries for every channel, asymptotic class, spin component, and frequency sector and indicate whether a data point belongs to the symmetry-reduced sector or -- otherwise -- how to retrieve a value via symmetry-relations. 

\subsubsection{Frequency grids}\label{sec:frequency_grid}

For numerical calculations, the continuous frequency dependence of correlation functions in the KF (and in the MF at $T=0$) must be discretized. Since these functions can become sharply peaked around certain frequencies, especially the lower the temperature, but simultaneously decay only slowly asymptotically (typically $\sim 1/\nu^2$ or even $\sim 1/\nu$ for some components), finding a suitable discretization that resolves all sharp structures but still captures the asymptotic decay is hard. Since the sharp features mostly occur at smaller frequencies (measured relative to the hybridization $\Delta$), we use a frequency grid that provides high resolution at small frequencies and fewer points at high frequencies. To achieve this, an equidistant grid of an auxiliary variable $\Omega\in [-1,1]$ is mapped to frequencies according to $\nu(\Omega)=A\Omega|\Omega|/\sqrt{1-\Omega^2}$. The parameter $A>0$ can be suitably chosen automatically or by hand for all quantities, as further explained in App.\,G of Ref.\,\onlinecite{main_paper}. However, we do not recommend optimizing $A$ automatically, as this can become expensive and unreliable in the presence of numerical artifacts.

The frequency grid is implemented in the \code{FrequencyGrid} class. It specifies the grid parameters such as the number of grid points or the scale factor $A$, and can access both continuous frequencies $\nu$ and auxiliary variables $\Omega$ corresponding to a given discrete index. Crucially, this also works the other way around, yielding the discrete index that corresponds to the frequency closest to a given continuous frequency. This is needed for interpolations, discussed in Sec.\,\ref{sec:interpolations}.

An instance of the \code{FrequencyGrid} class is instantiated in every instance of one of the correlation function classes, to parametrize their respective frequency dependencies. The vertex classes naturally require up to three instances of the \code{FrequencyGrid} each.

The frequency grids are rescaled during mfRG flow calculations which use the hybridization flow scheme, see Sec.\,\ref{sec:mfRG}. The \code{FrequencyGrid} class provides all functionality required for that purpose.

As a side note, two alternative frequency grids have been implemented. One is a hybrid grid which consists of a quadratic part at small frequencies, a linear part at intermediate frequencies, and a rational part at large frequencies. The other uses polar coordinates to parametrize the two-dimensional frequency dependence of three-point functions, i.e.\, the $K_2$ and $K_{2'}$ classes. Which grid is to be used is controlled by the \code{GRID} flag, see Sec.\,\ref{sec:parameters}. In our experience, the non-linear grid explained at the beginning of this section is the most useful if the scale parameters $A$ are chosen suitably.

\subsubsection{Frequency integration}\label{sec:integrations}

The following passage is taken almost verbatim from the Ph.D.\,thesis of E.\,Walter\cite{walter_keldyshfrg_2022}.\\

Computing numerical integrals with high accuracy is a crucial ingredient for obtaining correct results in the context of diagrammatic calculations discussed here. At the same time, the integrator is also critical for the performance of the computation, since evaluating integrals constitutes the computationally most expensive part of the code. For these reasons, we use an adaptive integration routine that automatically determines where to evaluate the integrand within the integration domain. Regions with sharp features require many evaluation points in order to get a high accuracy, while in regions where the integrand is smooth fewer evaluations suffice, which increases the performance of the computation. Such an adaptive integrator is really indispensable for the problem at hand: Non-adaptive routines like a simple trapezoidal or Simpson rule on an equidistant grid often lead to systematically wrong results.

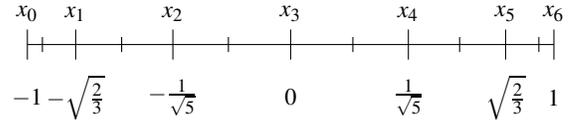
\begin{figure}
\begin{tikzpicture}
    \def\length{3.5};
   \draw (-\length,0) -- (\length,0);
   \draw (-\length,-0.2) -- (-\length,0.2);
   \draw (-0.942882*\length,-0.1) -- (-0.942882*\length,0.1);
   \draw (-0.816497*\length,-0.2) -- (-0.816497*\length,0.2);
   \draw (-0.641853*\length,-0.1) -- (-0.641853*\length,0.1);
   \draw (-0.447214*\length,-0.2) -- (-0.447214*\length,0.2);
   \draw (-0.236383*\length,-0.1) -- (-0.236383*\length,0.1);
   \draw (0,-0.2) -- (0,0.2);
   \draw (0.236383*\length,-0.1) -- (0.236383*\length,0.1);
   \draw (0.447214*\length,-0.2) -- (0.447214*\length,0.2);
   \draw (0.641853*\length,-0.1) -- (0.641853*\length,0.1);
   \draw (0.816497*\length,-0.2) -- (0.816497*\length,0.2);
   \draw (0.942882*\length,-0.1) -- (0.942882*\length,0.1);
   \draw (\length,-0.2) -- (\length,0.2);
   \node[above] at (-\length,0.2) {$x_0$};
   \node[above] at (-0.816497*\length,0.2) {$x_1$};
   \node[above] at (-0.447214*\length,0.2) {$x_2$};
   \node[above] at (0,0.2) {$x_3$};
   \node[above] at (0.447214*\length,0.2) {$x_4$};
   \node[above] at (0.816497*\length,0.2) {$x_5$};
   \node[above] at (\length,0.2) {$x_6$};
   \node at (-\length,-0.7) {$-1$};
   \node at (-0.816497*\length,-0.7) {$-\sqrt{\frac23}$};
   \node at (-0.447214*\length,-0.7) {$-\frac{1}{\sqrt{5}}$};
   \node at (0,-0.7) {$0$};
   \node at (0.447214*\length,-0.7) {$\frac{1}{\sqrt{5}}$};
   \node at (0.816497*\length,-0.7) {$\sqrt{\frac23}$};
   \node at (\length,-0.7) {$1$}; 
\end{tikzpicture}
\caption{Distribution of the nodes $x_j$ of the 4-point Gauss--Lobatto rule with 7-point and 13-point Kronrod extension. The lower row indicates the values of the nodes for integration boundaries $a=-1$, $b=1$.}
\label{fig:distribution}
\end{figure}

We use $n$-point integration rules that approximate integrals of the kind $\int_a^b F(x)\,\mathrm{d}x \approx \sum_{j=1}^n F(x_j)w_j$ with nodes $x_j$ and corresponding weights $w_j$.
The integrator we use and which is implemented in the \code{Adapt} class in the code is an adaptive 4-point Gauss--Lobatto routine with 7-point Kronrod extension and 13-point Kronrod extension as error estimate, as detailed in \onlinecite{press_numerical_2007}. The benefit of Gauss--Lobatto rules, compared to, e.g., the widely-used Gauss--Kronrod rules, is that the nodes include the endpoints of the integration domain. This allows us to subdivide the domain at the nodes of the integration rule and reuse points that have been computed previously, which is preferential in terms of performance. Similarly, the Kronrod extensions of a Gauss--Lobatto rule reuse all points from a corresponding lower-point rule and simply add additional points, which effectively allows us to get two different rules from one set of evaluation points.\\
The nodes $x_j$ of the 4-point Gauss--Lobatto rule with 7-point and 13-point Kronrod extension are distributed as shown in Fig.~\ref{fig:distribution}.
There, the lower row indicates the values of the nodes for integration boundaries $a=-1$, $b=1$ (for other values of $a$, $b$, the values have to be rescaled correspondingly).
The 4-point Gauss--Lobatto rule (GL4) and 4-point Gauss--Lobatto with 7-point Kronrod extension (GLK7) use the following points:
\begin{subequations}
\begin{align}
    \mathrm{GL4}(x_0,x_6) &= \sum_{j\in\{0,2,4,6\}} F(x_j) \, w_j\\
    \mathrm{GLK7}(x_0,x_6) &= \sum_{j=0}^6 F(x_j) \, w_j
    \,.
\end{align}
\end{subequations}
The smaller marks between the nodes $x_0,\dots,x_6$ in the graphical representation above indicate the additional 6 points that are added in the 13-point Kronrod extension (GLK13), which are only known numerically (these and the weights $w_j$ are found in \onlinecite{press_numerical_2007}).

\begin{figure}
    \begin{tikzpicture}
        \node at (0,0) {Error estimate: $I_s = \mathrm{GLK13}(a,b)$};
		\draw[->] (0,-0.4) -- (0,-0.8);
		\node at (0,-1.2) {$I_1=\mathrm{GL4}(a,b)\phantom{K}$};
		\node at (0,-1.6) {$I_2=\mathrm{GLK7}(a,b)$};
		\draw[->] (0,-2) -- (0,-2.4);
		\node at (0,-2.8) {$|I_2-I_1| < \epsilon \cdot I_s$};
		\draw (-1.4,-2.5) rectangle (1.4,-3.1);
		\draw[->] (1.5,-2.8) -- (2.5,-2.8);
		\node[above] at (1.8,-2.8) {yes};
		\node at (3.7,-2.8) {$\int_a^b F(x) \, \mathrm{d} x = I_2$};
		\draw[->] (0,-3.2) -- (0,-4.4);
		\node[right] at (0,-3.4) {no};
		\node[right] at (0,-4) {$x_0 \leftarrow a \,, \ x_6 \leftarrow b$};
		\node at (0,-4.8) {subdivide:};
		\node at (0,-5.4) {$[x_0,x_1], [x_1,x_2], \dots, [x_5,x_6]$};
		\draw (-2.3,-4.5) rectangle (2.3,-5.7);
		\draw[->] (-1.6,-5.8) -- (-1.6,-7);
		\draw[->,dotted] (-0.3,-5.8) -- (-0.3,-6.2);
		\draw[->,dotted] (1.52,-5.8) -- (1.52,-6.2);
		\node at (0.61,-6) {$\dots$};
		\node[right] at (-1.5,-6.6) {for each subinterval separately};
		\node at (-1,-7.4) {$I_1=\mathrm{GL4}(x_i,x_{i+1})\phantom{K}$};
		\node at (-1,-7.8) {$I_2=\mathrm{GLK7}(x_i,x_{i+1})$};
		\draw[->] (-1,-8.2) -- (-1,-8.6);
		\node at (-1,-9) {$|I_2-I_1| < \epsilon \cdot I_s$};
		\draw (-2.4,-8.7) rectangle (0.4,-9.3);
		\draw[->] (0.5,-9) -- (1.5,-9);
		\node[above] at (0.8,-9) {yes};
		\node at (3,-9) {$\int_{x_i}^{x_{i+1}} F(x) \, \mathrm{d} x = I_2$};
		\draw[->] (-1,-9.4) -- (-1,-9.8) -- (-3.2,-9.8) -- (-3.2,-5.1) -- (-2.4,-5.1);
		\node[right] at (-1,-9.6) {no};
		\node[below] at (-2.1,-9.9) {$x_0 \leftarrow x_i$};
		\node[below] at (-1.93,-10.3) {$x_6 \leftarrow x_{i+1}$};
    \end{tikzpicture}
    \caption{Schematic illustration of the integration algorithm; an adaptive 4-point Gauss--Lobatto routine with 7-point Kronrod extension and 13-point Kronrod extension as an error estimate.}
    \label{fig:integrator}
\end{figure}
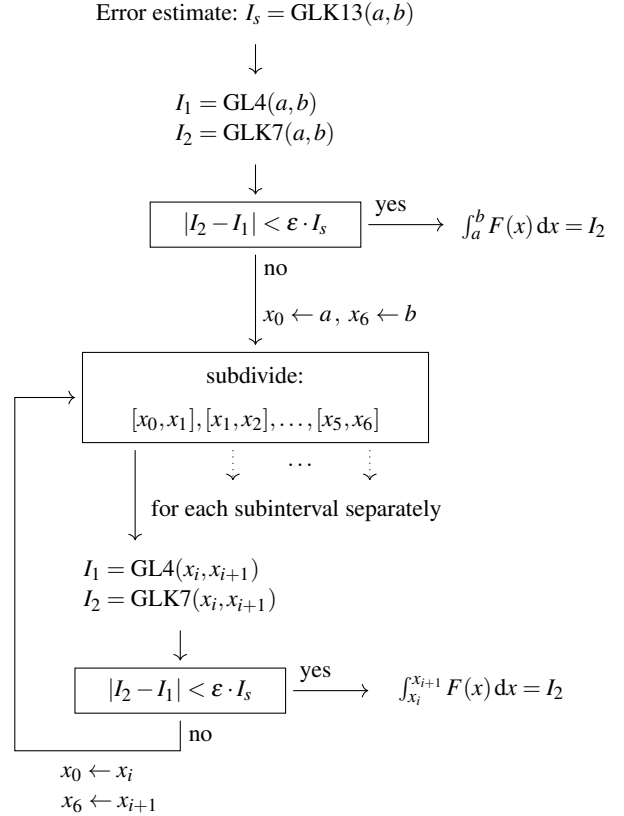

The recursive algorithm of the integrator then works as shown in Fig.~\ref{fig:integrator}.
Note that the error estimate $I_s$ is determined only once for the full integral and then reused for each subinterval, in order to avoid infinite recursions in subintervals.
A typical recommended value the relative accuracy is $\epsilon=10^{-5}$, which is set by the global variable \code{integrator\_tol}, see Tab.\,\ref{tab:global_parameters}.

\subsubsection{Asymptotic corrections to frequency integrals}

In the previous section it was explained how frequency integrations over a finite interval $[a,b]$ are performed. Since diagrammatic calculations require integrations over the full frequency axis (or summations over an infinite set of discrete Matsubara frequencies for the finite-$T$ MF), the contributions to the integral resulting from the high-frequency asymptotics of the integrands have to be treated as well. This is particularly relevant for slowly decaying integrands, which occur often, as the correlation functions arising in the present context typically only decay as $\sim 1/\nu$ or $\sim 1/\nu^2$.

In the KF and the zero-$T$ MF, involving continuous frequency integrations, a naive treatment turned out to be sufficient: Since the frequency axes are discretized non-uniformly, as described in Sec.\,\ref{sec:frequency_grid}, the largest discrete frequency grid point is always so large that the high-frequency tails can be treated via quadrature, ignoring the minuscule contributions of even larger frequencies. For finite-$T$ MF computations, which involve infinite sums, the code provides two options for the treatment of high-frequency tails in the integrand: (i) The tails can be treated via quadrature, by approximating the sum with an integral and then following the same logic as in the KF. (ii) For bubble computations, the lowest order contribution from the bare bubble, which is known analytically, can be used. This is justified by the fact that in the high-frequency asymptotic limit, the non-trivial contributions due to interactions, encoded in the self-energy, have decayed and only the bare contribution is responsible for the asymptotic behavior. The first or second option is chosen with the \code{ANALYTIC\_TAILS} parameter flag, see also Sec.\,\ref{sec:parameters}

\subsubsection{Interpolation routines}\label{sec:interpolations}

Whenever the value of a correlation function at some continuous frequency argument is required, in particular during frequency integrations, the data stored on discrete frequency grids has to be interpolated. In addition, the diagrammatic algorithms discussed here have feedback between the three two-particle channels, which all have their own channel-dependent parametrizations. This necessitates accurate interpolations between different frequency parametrizations, otherwise errors accumulate over the course of a computation.

To handle the interpolation of multidimensional correlation functions, we implemented multilinear interpolation and cubic spline interpolation using cubic Hermite splines. While spline interpolation is robust against minor inaccuracies of the data points and offers faster convergence in the number of frequency points for smooth functions, multilinear interpolation is generally faster numerically. Having tried out both options, we prefer linear interpolation, as spline interpolation only really becomes useful for better precision if the function is already well resolved.

Regarding linear interpolation, the code offers options: One can either interpolate on the grid of frequencies $\nu$, or on the grid of auxiliary frequencies $\Omega$, which are equidistantly spaced on the interval $[-1,1]$, see Sec.\,\ref{sec:frequency_grid}. We found the latter option to be more accurate. The global parameter \code{INTERPOLATION} specifies which type of interpolation shall be used, see also Sec.\,\ref{sec:parameters}.

\subsubsection{Data structures}\label{sec:data_structures}

The central low-level data structure used for storing and retrieving numerical data inside the code is the \code{dataBuffer} class. It was devised with the two main intentions of efficiency and flexibility in mind, see also our discussion of the main design choices for the codebase in the introduction, Sec.\,\ref{sec:introduction}. On the one hand, it should enable building integrands that return scalar- or vector-valued entries as efficiently as possible, particularly avoiding conditional (``if-else'') statements during runtime, as these prevent optimizations such as loop-vectorization or function inlining. On the other hand, it should be usable in all parts of the codebase, e.g., for both calculations with interpolations on continuous frequency grids and for finite-$T$ MF calculations, which only require indexing of discrete data points.

The \code{dataBuffer} class is structured as follows. It builds upon the \code{dataContainerBase} class, which is used to represent multi-dimensional tensors, allowing scalar and vector-valued access to contiguous elements. The \code{DataContainer} class then inherits \code{dataContainerBase}, adding frequency information. It contains a multi-dimensional frequency grid (see Sec.\,\ref{sec:frequency_grid}) to parametrize all its associated frequency arguments and provides functions to analyze the resolution of frequency grids. Inheriting the \code{DataContainer} class, the \code{Interpolator} classes then implement the different interpolation routines outlined in Sec.\,\ref{sec:interpolations}. Multilinear cubic spline interpolations require pre-computation and storage of interpolation coefficients, whereas linear interpolations happen on the fly. Finally, the \code{dataBuffer} class inherits both the \code{Interpolator} and the \code{DataContainer} classes and can be used in actual computation. In addition, it can update and optimize grid parameters as required.

\subsubsection{Template arguments}

Another performance-critical aspect of the codebase is its heavy use of templates. In particular, the propagation of template arguments as specified by preprocessor flags enables the determination of the required diagrammatic combinations for any given computation at compile time. Selecting and combining the necessary vertex contributions this way, e.g., for contributions to specific asymptotic classes, enables further optimization by the compiler. However, the ubiquity of template arguments comes at the expense of readability in many places.

\subsection{Tests}\label{sec:testing}

The code includes a large number ($178$ as of writing) of self-explanatory unit tests that run checks on the low-level parts of the codebase. They are implemented using the popular \code{Catch2} library\cite{catch2} and are invoked from a separate C++ source file \code{unit\_tests.cpp}, which should be built separately from the main source file. From inside this file, more involved and expensive tests can be started if desired. These include detailed tests of the ODE solver or perturbation theory, which are too expensive to be part of the unit test suite. Lastly, the code includes functionality to produce reference data that can be used later to compare the results of a calculation after changes to the code have been made. We have found it immensely useful to include many unit tests into the codebase, as they can tell almost immediately if a single technical part of the code has broken. Moreover, having a way to compare the results of very involved computations that involve large parts of the codebase at once is useful to catch logical errors. We wholeheartedly recommend both.

\subsection{Parameters}\label{sec:parameters}

\begin{table*}[]
    \centering
    \begin{tabular}{l l p{70 ex}}
        \toprule
        macro name  & possible values & description \\ 
        \midrule
        \code{ADAPTIVE\_GRID} & - & if defined, use optimization routine to find the best scale factor A of the frequency grid; if undefined, just rescale the grid. Warning: Can be expensive and unreliable in the presence of numerical artifacts. \\
        \code{ANALYTIC\_TAILS} & 0, 1 & 0 for false; 1 for true. If true, the analytic expression for the bare bubble is used to treat the high-frequency asymptotics during bubble computations in the finite-$T$ MF. \\
        \code{BARE\_SE\_FEEDBACK} & - & If defined, only bare selfenergy is used. Only makes sense if \code{STATIC\_FEEDBACK} is defined. Useful for benchmarks with previous Keldysh fRG schemes. \\
        \code{CONTOUR\_BASIS} & 0, 1 & 0 for false, 1 for true: If true, no Keldysh rotation is performed and the contour basis is used instead to parametrize the Keldysh components of all correlation functions. Useful for comparisons with results that use this convention. Not as well tested and thus not recommended for production runs. \\
        \code{DEBUG\_SYMMETRIES} & 0, 1 & 0 for false; 1 for true. Performs computations without use of symmetries if true. Useful for debugging purposes. \\
        \code{GRID} & 0, 1, 2 & Controls which frequency grid is to be used. 0 for the non-linear grid, 1 for the hybrid grid, 2 for the polar grid. Recommendation: 0. See also Sec.\,\ref{sec:frequency_grid}. \\
        \code{KATANIN} & - & If defined, the Katanin extension is used during fRG computations. \\
        \code{KELDYSH\_FORMALISM} & & Determines whether calculations shall be done in the Keldysh or Matsubara formalism. 0 for Matsubara formalism (MF); 1 for Keldysh formalism (KF). \\
        \code{MAX\_DIAG\_CLASS} & 1, 2, 3 & Defines the diagrammatic classes that will be considered: 1 for only $K_1$, 2 for $K_1$ and $K_2$ and 3 for the full dependencies. Useful for debugging purposes and for computations in second-order perturbation theory, or if \code{STATIC\_FEEDBACK} is defined, when only $K_1$ is required. \\
        \code{NDEBUG}           & -- & If defined, assert functions are switched off. Recommended setting for production runs. \\
        \code{PARTICLE\_HOLE\_SYMM} & 0, 1 & 0 for false; 1 for true. If true, particle-hole symmetry is assumed. \\
        \code{PT2\_FLOW} & - & If defined, only compute the flow equations up to $O(U^2)$. Only makes sense for pure $K_1$ calculations. Useful as a consistency check together with independent PT2 calculations. \\
        \code{REG} & 2, 3, 4, 5 & Specifies the mfRG flow regulator to be used. $2$: $\Delta$-flow, $3$: $\omega$-flow, $4$: $U$-flow, $5$: $T$-flow. For details, see Sec.\,\ref{sec:flow_schemes}. \\
        \code{REPARAMETRIZE\_FLOWGRID} & - & If defined, the flow parameter is reparametrized according to Sec.\,\ref{sec:ODE-details}. Only recommended for the $\Delta$-flow. \\
        \code{SBE\_DECOMPOSITION} & 0, 1 & 0 for false; 1 for true. If true, the SBE decomposition is used to parametrize the vertex and the flow equations. Only implemented in the MF!  \\
        \code{SELF\_ENERGY\_FLOW\_CORRECTIONS} & 0, 1 &  0 for false; 1 for true. If true, corrections to the flow equations for the vertex from the self-energy, starting at $\ell=3$, are included. \\
        \code{STATIC\_FEEDBACK} & - & If defined, use static $K_1$ inter-channel feedback as done in \onlinecite{jakobs_functional_2010}. Only makes sense for pure $K_1$ calculations. \\
        \code{SWITCH\_SUM\_N\_INTEGRAL} & 0, 1 & 0 for false; 1 for true. If true, the sum over internal Keldysh indices is done before the frequency integration. Recommended setting: 1. \\
        \code{USE\_ANDERSON\_ACCELERATION} & 0, 1 &  0 for false; 1 for true. If true, Anderson acceleration is used to converge parquet iterations and self-energy iterations in mfRG faster. \\
        \code{USE\_MPI} & - & If defined, MPI is used for parallelization across multiple nodes. \\
        \code{USE\_SBEb\_MFRG\_EQS} & 0, 1 & Determines which version of the SBE approximation shall be used. 0 for SBEa, 1 for SBEb. Only implemented in the MF!\\
        \code{VECTORIZED\_INTEGRATION} & 0, 1 & 0 for false; 1 for true. If true, integrals are performed with vector-valued integrands. For Keldysh, vectorization over Keldysh indices. For Matsubara at finite $T$, vectorization over the Matsubara sum. \\
        \code{ZERO\_TEMP} & 0, 1 & 0 for false; 1 for true. If true, temperature $T = 0$ is assumed. \\
        \bottomrule
    \end{tabular}
    \caption{Incomplete list of the most important preprocessor macros, to be set before compilation.}
    \label{tab:preprocessor_macros}
\end{table*}

\begin{table*}[]
    \centering
    \begin{tabular}{l l p{60 ex}}
        \toprule
        parameter name  & type & description \\ 
        \midrule
        \code{converged\_tol} & \code{double} & Tolerance for loop convergence in mfRG. \\
        \code{COUNT} & \code{int} & Used to set the number of frequency points in the MF. For details, see the definitions in the file \code{frequency\_parameters.hpp}.\\
        \code{Delta\_factor\_K1} & \code{int} & Scale factor for the frequency grid of the K1 vertex class.\\
        \code{Delta\_factor\_SE} & \code{int} & Scale factor for the frequency grid of the self-energy.\\
        \code{Delta\_factor\_K2\_w} & \code{int} & Scale factor for the frequency grid of the bosonic frequency of the K2 and K2' vertex classes.\\
        \code{Delta\_factor\_K2\_v} & \code{int} & Scale factor for the frequency grid of the fermionic frequency of the K2 and K2' vertex classes.\\
        \code{Delta\_factor\_K3\_w} & \code{int} & Scale factor for the frequency grid of the bosonic frequency of the K3 vertex class.\\
        \code{Delta\_factor\_K3\_v} & \code{int} & Scale factor for the frequency grid of the fermionic frequencies of the K3 vertex class.\\
        \code{EQUILIBRIUM} & \code{bool} & If \code{true}, use equilibrium FDRs for propagators. \\
        \code{glb\_mu} &  \code{double} & Chemical potential -- w.l.o.g. ALWAYS set to 0.0 for the AM! \\
        \code{integrator\_tol} & \code{double} & Integrator tolerance. \\
        \code{inter\_tol} & \code{double} & Tolerance for closeness to grid points when interpolating. \\
        \code{INTERPOLATION} & \code{linear}, \code{linear\_on\_aux}, \code{cubic} & Interpolation method to me used. \code{linear}: linear interpolation on the frequency grid. \code{linear\_on\_aux}: linear interpolation on the grid for the auxiliary frequency $\Omega$. \code{cubic}: Interpolation with cubic splines (warning: expensive!). \\
        \code{Lambda\_ini} &  \code{double} & Initial value of the regulator $\Lambda$ for an mfRG flow. \\
        \code{Lambda\_fin} &  \code{double} & Final value of the regulator $\Lambda$ for an mfRG flow. \\
        \code{Lambda\_scale} & \code{double} & Scale of the log substitution, relevant in the hybridization flow. \\
        \code{dLambda\_initial} & \code{double} & Initial step size for ODE solvers with adaptive step size control. \\
        \code{nBOS} & \code{int} & Number of bosonic frequency points for the $K_1$ vertex class. \\
        \code{nFER} & \code{int} & Number of fermionic frequency points for the self-energy. \\
        \code{nBOS2} & \code{int} & Number of bosonic frequency points for the $K_2$ and $K_{2'}$ vertex classes. \\
        \code{nFER2} & \code{int} & Number of fermionic frequency points for the$K_2$ and $K_{2'}$ vertex classes. \\
        \code{nBOS3} & \code{int} &  Number of bosonic frequency points for the $K_3$ vertex class. \\
        \code{nFER3} & \code{int} & Number of fermionic frequency points for the $K_3$ vertex class. \\
        \code{U\_NRG} & \code{std::vector<double>} & Vector with the values of $U$ in units of $\Delta$ that an mfRG flow should cover. Serve as checkpoints for the flow. Useful for benchmarking purposes if data from other methods at precise parameter points are available. \\
        \code{VERBOSE}   & \code{bool}  & If \code{true}, detailed information about all computational steps is written into the log file. Recommended setting for production runs: \code{false} \\
        \code{nmax\_Selfenergy\_iterations} & \code{int} & Maximal number of self-energy iterations to be done during an mfRG flow for $\ell \geq 3$. Default value: 10\\
        \code{tol\_selfenergy\_correction\_abs} & \code{double} & Absolute tolerance for self-energy iterations in mfRG. Default value: $10^{-9}$ \\
        \code{tol\_selfenergy\_correction\_rel} & \code{double} & Relative tolerance for self-energy iterations in mfRG. Default value: $10^{-5}$ \\
        \bottomrule
    \end{tabular}
    \caption{Incomplete list of global parameters, to be set before compilation.}
    \label{tab:global_parameters}
\end{table*}

Before any individual calculation can be started, a number of parameters have to be set. As the code provides a large degree of flexibility, the number of possible parameter choices is large. Most of these parameters are set inside the corresponding header files before compilation. The reason for this is that, depending on these choices, often different functionality of the code is invoked, depending, e.g., on the choice of formalism. This is achieved by defining preprocessor macros accordingly, which makes the corresponding functionality accessible. As discussed previously in Sec.\,\ref{par:flags}, while this approach was useful for implementing new functionality quickly, in the long run, it turned out to be problematic with regard to readability and maintainability of the code. Table \ref{tab:preprocessor_macros} provides a list (albeit incomplete) of the most important preprocessor flags used in the code with a short description of each. 

In addition, global parameters have to be set, which specify settings like the resolution of the frequency grid, convergence criteria, or start- and end-points of an mfRG flow. Table \ref{tab:global_parameters} provides a non-exhaustive list of those.

Lastly, it should be mentioned that once the code has been compiled and the resulting executable is to be called, it requires three run-time arguments: The first one invokes an mfRG run if it is a positive integer, specifying the maximal number of loop orders calculated during the mfRG flow. Alternatively, if it is set to $0$ or $-1$, a parquet or PT2 calculation are started, respectively. The second is a positive integer and specifies the number of nodes to be utilized. The third runtime argument defines the temperature for the calculation and was introduced to easily enable parameter sweeps without having to recompile the code every time. Note that its value is irrelevant for calculations that have the flag \code{ZERO\_TEMP} set to $1$ or if an mfRG run is performed with the flag \code{REG} set to $5$, which employs the temperature flow.

\section{Algorithms}\label{sec:algorithms}

In the third main part of the paper, we finally describe three diagrammatic algorithms that have been implemented. These are second-order perturbation theory (PT), a self-consistent solution of the parquet equations, and the flow equations provided by the multiloop functional renormalization group (mfRG). For all three methods, we first give some theoretical background, before describing schematically how the algorithms are implemented and what functions are being used.

\subsection{Perturbation theory}\label{sec:perturbation_theory}

The simplest computations that can be done with the code are perturbation theory calculations. While these are easy to implement in the second order, going to higher orders involves an increasing number of diagrams which can in principle be evaluated separately. This is, however, not always straightforward, e.g\,if symmetries are to be exploited: individual diagrams of the perturbation series do not all have the same symmetries as a full vertex, such that symmetry-related diagrams have to be provided, which can become tedious. Alternatively, the flag \code{DEBUG\_SYMMETRIES} can be set to 1, see Sec.\,\ref{sec:parameters}, in which case the code does not attempt to exploit symmetries. As higher-order perturbation theory has so far only been done for testing purposes and consistency checks (see, e.g., Chapter 7 in \onlinecite{walter_keldyshfrg_2022}), we refrain from going into further detail here. Instead, we focus just on the second-order case, and on Hartree-Fock theory for the self-energy, relevant to the aAM.

\subsubsection{Hartree-Fock}\label{sec:Hartree}

As elaborated in Ref.\,\onlinecite{main_paper}, it is helpful to replace the bare propagator $G_0$ by the Hartree-propagator $G_\mathrm{H}$, which is shifted by the Hartree-term of the self-energy,
\begin{align}
G^R_0 \to G^R_\mathrm{H} = \frac{1}{\nu - \epsilon_d + i\Delta - \Sigma^R_\mathrm{H}}
.\label{eq:Hartree-propagator}
\end{align}
For the sAM, this is almost trivial, as the retarded component of the Hartree term reads $\Sigma^R_\mathrm{H}=U/2$, which simply yields $G^R_\mathrm{H}= (\nu + i\Delta)^{-1}$. For the aAM, on the other hand, the Hartree-term can be computed self-consistently.

For this purpose, the class \code{Hartree\_Solver} provides the function \code{compute\_Hartree\_term\_bracketing}. It computes $\Sigma^R_\mathrm{H}$ via
\begin{align}
    \Sigma_\mathrm{H}^R = U \int \frac{\mathrm{d} \nu'}{2\pi i} \ G^<_\mathrm{H} (\nu'),
    \label{eq:Hartree}
\end{align}
where in thermal equilibrium, the relation $G^< (\nu) = -2i\, n_F(\nu) \mathrm{Im}\, G^R (\nu)$ is used with the Fermi function $n_F(\nu) \!=\! 1/(1 \!+\! e^{\nu/T})$. As $\Sigma_\mathrm{H}^R$ enters both sides of Eq.\,\eqref{eq:Hartree}, this calculation is done self-consistently, using a simple bracketing algorithm. 

In addition, the \code{Hartree\_Solver} class provides the function \code{compute\_Hartree\_term\_oneshot}, which evaluates Eq.\,\eqref{eq:Hartree} just once, given a provided self-energy for $G^R (\nu)$. This function is invoked in the context of parquet iterations and evaluations of mfRG flow equations to update the Hartree term of the aAM. 

Lastly, the \code{Hartree\_solver} class provides functionality to check the fulfillment of the Friedel sum rule \cite{Langreth1966} $\langle n_\sigma \rangle = \tfrac{1}{2} \!-\! \tfrac{1}{\pi} \arctan [(\epsilon_d+\Sigma(0))/\Delta]$, which the self-consistent Hartree term fulfills at $T \!=\! 0$.

\subsubsection{Second order perturbation theory (PT2)}\label{sec:PT2}

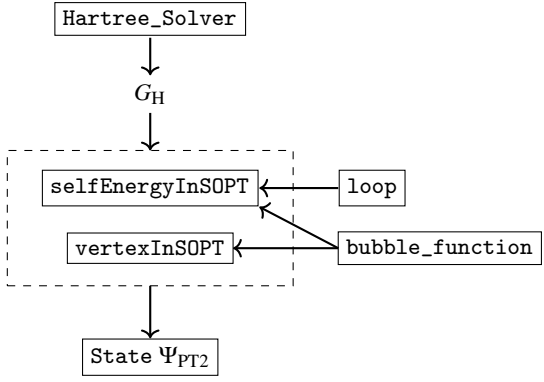
\begin{figure}
    \begin{tikzpicture}
        \node[draw] at (0,0) {\code{Hartree\_Solver}};
        \node at (0, -1) {$G_\mathrm{H}$};
        \node[draw] at (0,-2.25) {\code{selfEnergyInSOPT}};
        \node[draw] at (0,-3.05) {\code{vertexInSOPT}};
        \draw[dashed] (-1.9, -3.55) rectangle (1.9, -1.75);
        \node[draw] at (0, -4.5) {\code{State} $\Psi_\mathrm{PT2}$};
        \node[draw] at (2.95, -2.25) {\code{loop}};
        \node[draw] at (3.85, -3.05) {\code{bubble\_function}};
        \draw[->, thick] (0, -0.25) -- (0, -0.75);
        \draw[->, thick] (0, -1.25) -- (0, -1.75);
        \draw[->, thick] (0, -3.55) -- (0, -4.25);
        \draw[->, thick] (2.5, -2.25) -- (1.45, -2.25);
        \draw[->, thick] (2.5, -3.05) -- (1.1, -3.05);
        \draw[->, thick] (2.5, -3.05) -- (1.45, -2.5);
    \end{tikzpicture}
    \caption{Schematic depiction of the function \code{sopt\_state}.}
    \label{fig:sopt}
\end{figure}

The self-energy and vertex in second-order perturbation theory are computed via the function \code{sopt\_state}, which works as depicted in Fig.~\ref{fig:sopt}.
It first initializes a bare state see Sec.\,\ref{sec:state}, given the system parameters and the current value of the regulator $\Lambda$. For the aAM, this already includes a self-consistent calculation of the Hartree term, see Sec.\,\ref{sec:Hartree}. Then, it invokes the function \code{selfEnergyInSOPT}, which computes the single diagram for the dynamical part of the self-energy in PT2 by first computing a bare bubble in the $a$-channel using the \code{bubble\_function}, see Sec.\,\ref{sec:bubble_function}, with two bare vertices, and then closing the loop over that bare bubble with the Hartree-propagator using the \code{loop} function, see Sec.\,\ref{sec:loop}.

Thereafter, the vertex is computed using the function \code{vertexInSOPT}, which simply invokes the \code{bubble\_function} three times, once for each of the three two-particle channels $a$, $p$ and $t$, using two bare vertices, adding each result to the vertex.

In total, this procedure yields all diagrams for the dynamical part of the self-energy and the vertex in PT2, using the Hartree-propagator $G_\mathrm{H}$. For the precise diagrammatic definition of PT2, see App.\,F in Ref.\,\onlinecite{main_paper}.

\subsection{Parquet equations}\label{sec:parquet}

The parquet formalism\cite{bickers_self-consistent_2004} provides a self-consistent set of equations for the self-energy $\Sigma$ and the three two-particle reducible vertices $\gamma_r$ with $r \in \{a, p, t\}$. The latter are given by the Bethe-Salpeter equations (BSEs), 
\begin{subequations}\label{eq:BSE}
\begin{align}
    \tikzm{gamma_a}{
        \fullvertexwithlegs{$\gamma_a$}{0}{0}{1}
    }
    &=
    \tikzm{mfRG-parquet-BSE_a1}{
			\arrowslefthalffull{0}{0}{1}
			\fullvertex{$I_a$}{0}{0}{1}
			\abubblefullfull{0.3}{0}{1}{1}{1}
			\fullvertex{$\Gamma$}{1.8}{0}{1}
			\arrowsrighthalffull{1.8}{0}{1}
		}
    ,
    \\
    \tikzm{gamma_p}{
        \fullvertexwithlegs{$\gamma_p$}{0}{0}{1}
    }
    &=
    \frac{1}{2} \!\!
    \tikzm{mfRG-parquet-BSE_p1}{
			\arrowslefthalffullp{0}{0}{1}{1}{1}
			\fullvertex{$I_p$}{0}{0}{1}
			\pbubblefullfull{0.3}{0}{1}{1}{1}
			\fullvertex{$\Gamma$}{1.8}{0}{1}
			\arrowsrighthalffullp{1.8}{0}{1}{1}{1}
		}
    ,
    \\
    \tikzm{gamma_t}{
        \fullvertexwithlegs{$\gamma_t$}{0}{0}{1}
    }
    &=
      \ - \ \
    \tikzm{mfRG-parquet-BSE_1}{
			\arrowsupperhalffull{0}{0.75}{1}
			\fullvertex{$I_t$}{0}{0.75}{1}
			\tbubblefullfull{0}{0.45}{1}{1}{1}
			\fullvertex{$\Gamma$}{0}{-0.75}{1}
			\arrowslowerhalffull{0}{-0.75}{1}
		} \, 
    ,
\end{align}
\end{subequations}
where $I_r = \Gamma - \gamma_r$ is the two-particle irreducible vertex in channel $r$.
The self-energy is given by the Schwinger-Dyson equation (SDE),
\begin{align}\label{eq:SDE}
    \tikzm{selfenergy}{
        \selfenergywithlegs{$\Sigma$}{0}{0}{1}
    }
    =
    -
    \tikzm{mfRG-parquet-SDE0}{
			\barevertex{0}{0}
			\loopbarevertex{lineWithArrowCenter}{0}{0}
			\arrowslowerhalf{0}{0}
			\node at (0,0.45) {};
		}
    -
    \frac{1}{2} \
    \tikzm{mfRG-parquet-SDE_Gamma}{
			\barevertex{0}{0}
			\abubblebarefull{0}{0}{1}{1}
			\fullvertex{$\Gamma$}{1.5}{0}{1}
			\draw[lineWithArrowCenterEnd] (0,0) -- (-0.3,-0.3);
			\draw[lineWithArrowCenterEnd] (2.1,-0.6) -- (1.8,-0.3);
			\draw[lineWithArrowCenter] (1.8,0.3) .. controls ++(45:0.5) and ++(0:0.5) .. (0.9,0.8) .. controls ++(180:0.5) and ++(135:0.5) .. (0,0);
			\node at (0,0.85) {};
		}
    ,
\end{align}
which includes the Hartree term discussed in Sec.\,\ref{sec:Hartree}. Together, these equations close once the fully irreducible vertex $R$ is provided, for example by employing the PA, as discussed in Sec.\,\ref{sec:methods}.

\begin{figure}
\vspace{2ex}
    \begin{tikzpicture}
        \node at (0,0) {\boxed{\code{State}}};
        \node at (2.5,0.4) {\boxed{\code{compute\_BSE}}};
        \node at (2.5,-0.4) {\boxed{\code{compute\_SDE}}};
        \node at (5.85,0.4) {\boxed{\code{bubble\_function}}};
        \node at (4.95,-0.4) {\boxed{\code{loop}}};
        \node at (5.85,-1.2) {\boxed{\code{Hartree\_Solver}}};
        \draw[dashed] (1.1, -0.9) rectangle (3.9, 0.9);
        \draw[->, thick] (4.5, 0.4) -- (3.55, 0.4);
        \draw[->, thick] (4.5, -0.4) -- (3.55, -0.4);
        \draw[->, thick] (4.5, 0.4) -- (3.55, -0.2);
        \draw[->, thick] (4.55, -1.2) -- (3.55, -0.6);
        \draw[->, thick] (0,0.225) -- (0, 1.3) -- (2.5, 1.3) -- (2.5, 0.9);
        \draw[->, thick] (2.5, -0.9) -- (2.5, -1.3) -- (0, -1.3) -- (0, -0.22);
    \end{tikzpicture}
    \caption{Schematic depiction of the \code{parquet\_solver} function.}
    \label{fig:parquet}
\end{figure}
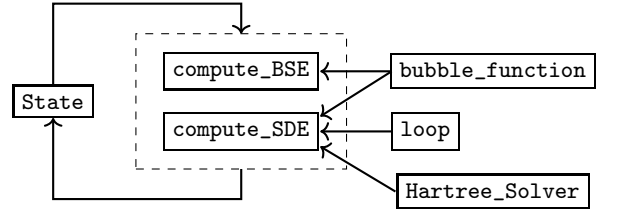

In practice, these equations are solved iteratively. The code provides functions to evaluate the right-hand sides of the BSEs and the SDE, called \code{compute\_BSE} and \code{compute\_SDE}. Schematically, the parquet solver works as depicted in Fig.~\ref{fig:parquet}.
Inside the code, a parquet computation is started by the function \code{run\_parquet}. It first initializes a state using PT2, as detailed in Sec.\,\ref{sec:PT2} before the \code{parquet\_solver} function is called. Internally, the \code{parquet\_solver} calls \code{parquet\_iteration}, which evaluates the BSEs and the SDE, given a provided input state, and combines them into an output state. The corresponding functions \code{compute\_BSE} and \code{compute\_SDE} use the machinery described in Sec.\,\ref{sec:objects} and Sec.\,\ref{sec:functionality} to evaluate Eqs.\,\eqref{eq:BSE} and \eqref{eq:SDE}. In practice, symmetrizing Eqs.\,\eqref{eq:BSE}, i.e.\, computing the sum of the right-hand side as is and with $I_r$ and $\Gamma$ interchanged and dividing by two, has proven beneficial for stability. Also, we found it helpful to combine all three ways to evaluate the SDE, Eq.\,\eqref{eq:SDE}, see App.\,D in Ref.\,\onlinecite{main_paper}.

The \code{parquet\_solver} can either proceed directly from one iteration to the next, or it can combine multiple results from previous iterations using mixing schemes to improve convergence. For example, one can combine the two most recent iterations with a mixing factor as outlined in Eq.\,(G4) of Ref.\,\onlinecite{main_paper}. One may start with a mixing factor of around $0.5$, which can be reduced automatically if the convergence properties of the calculation are poor. In addition, one can use Anderson acceleration \cite{anderson_iterative_1965, walker_anderson_2011} to combine multiple previous iterations for a prediction of the next iteration. We have found that this leads to faster convergence in the vicinity of the solution, but does not extend the parameter range where convergence can be reached.

\subsection{mfRG}\label{sec:mfRG}

In fRG \cite{RevModPhys.84.299}, the self-energy and vertex are interpolated between the initial and final values of a single-particle parameter $\Lambda$ introduced into the bare propagator $G_0$. The initial value $\Lambda=\Lambda_i$ should be chosen such that the theory is solvable at that point; in practice, it typically suffices that very good approximations of $\Sigma^{\Lambda_i}$ and $\Gamma^{\Lambda_i}$ can be obtained by PT2 or by converging the parquet equations.
The fRG then provides a set of differential ``flow'' equations in $\Lambda$ for $\Sigma^\Lambda$ and $\Gamma^\Lambda$ which yield the final results $\Sigma^{\Lambda_f}$ and $\Gamma^{\Lambda_f}$ at the actual point of interest $\Lambda = \Lambda_f$. In the multiloop fRG framework, these flow equations are derived from the parquet equations by differentiation with respect to the flow parameter $\Lambda$, as detailed in Ref.\, \onlinecite{kugler_derivation_2018}. This yields an infinite set of contributions of increasing ``loop order'' $\ell$,
\begin{subequations}
\begin{align}
    \dot{\Gamma} = \sum_{r\in\{a,p,t\}} \dot{\gamma}_r \\
    \dot{\gamma}_r = \sum_{\ell=1}^\infty \dot{\gamma}_r^{(l)} ,
\end{align}
\end{subequations}
where a dot represents a derivative w.r.t.\,$\Lambda$. Diagrammatically, the $\ell$-loop contributions in the $a$ channel read 
\begin{widetext}
\begin{subequations}
\label{eq:mfRG:mfRG:Flow_equations}
\begin{align}
	\tikzm{mfRG-ml-dgamma1}{
		\fullvertexwithlegs{$\dot{\gamma}_a^{(1)}$}{0}{0}{1.2}
	}
	&=
	\tikzm{mfRG-ml-dgamma1_rhs}{
		\arrowslefthalffull{0}{0}{1}
		\fullvertex{$\Gamma$}{0}{0}{1}
		\abubblefullfulldiff{0.3}{0}{1}{1}{1}
		\fullvertex{$\Gamma$}{1.8}{0}{1}
		\arrowsrighthalffull{1.8}{0}{1}
	}
	\,,
	\label{eq:mfRG:mfRG:Flow_equations:1l}
    \\
    \tikzm{mfRG-ml-dgamma2}{
		\fullvertexwithlegs{$\dot{\gamma}_a^{(2)}$}{0}{0}{1.2}
	}
	&= 
	\tikzm{mfRG-ml-dgamma2_rhs_L}{
		\arrowslefthalffull{0}{0}{1.2}
		\fullvertex{$\dot{\gamma}_{\bar{a}}^{(1)}$}{0}{0}{1.2}
		\abubblefullfull{0.36}{0}{1}{1.2}{1}
		\fullvertex{$\Gamma$}{1.86}{0}{1}
		\arrowsrighthalffull{1.86}{0}{1}
		\tikzunderbrace{$\dot{\gamma}_{a,L}^{(2)}$}{-0.66}{2.46}{-1}{-0.6}
	}+
	\tikzm{mfRG-ml-dgamma2_rhs_R}{
		\arrowslefthalffull{0}{0}{1}
		\fullvertex{$\Gamma$}{0}{0}{1}
		\abubblefullfull{0.3}{0}{1}{1}{1.2}
		\fullvertex{$\dot{\gamma}_{\bar{a}}^{(1)}$}{1.86}{0}{1.2}
		\arrowsrighthalffull{1.86}{0}{1.2}
		\tikzunderbrace{$\dot{\gamma}_{a,R}^{(2)}$}{-0.6}{2.52}{-1}{-0.6}
	}
	\,,
	\label{eq:mfRG:mfRG:Flow_equations:2l}
    \\
    \tikzm{mfRG-ml-dgamma3}{
		\arrowslefthalffull{0}{0}{1.2}
		\fullvertexwide{$\dot{\gamma}_a^{(\ell+2)}$}{0}{0}{1.2}{0.3}
		\arrowsrighthalffull{0.3}{0}{1.2}
	}
	&=
	\tikzm{mfRG-ml-dgamma3_rhs_L}{
		\arrowslefthalffull{0}{0}{1.2}
		\fullvertexwide{$\dot{\gamma}_{\bar{a}}^{(\ell+1)}$}{0}{0}{1.2}{0.3}
		\abubblefullfull{0.66}{0}{1}{1.2}{1}
		\fullvertex{$\Gamma$}{2.16}{0}{1}
		\arrowsrighthalffull{2.16}{0}{1}
		\tikzunderbrace{$\dot{\gamma}_{a,L}^{(\ell+2)}$}{-0.66}{2.76}{-1}{-0.6}
	}
	+
	\tikzm{mfRG-ml-dgamma3_rhs_C}{
		\arrowslefthalffull{0}{0}{1}
		\fullvertex{$\Gamma$}{0}{0}{1}
		\abubblefullfull{0.3}{0}{1}{1}{1.2}
		\fullvertex{$\dot{\gamma}_{\bar{a}}^{(\ell)}$}{1.86}{0}{1.2}
		\abubblefullfull{2.22}{0}{1}{1.2}{1}
		\fullvertex{$\Gamma$}{3.72}{0}{1}
		\arrowsrighthalffull{3.72}{0}{1}
		\tikzunderbrace{$\dot{\gamma}_{a,C}^{(\ell+2)}$}{-0.6}{4.32}{-1}{-0.6}
	}
	+
	\tikzm{mfRG-ml-dgamma3_rhs_R}{
		\arrowslefthalffull{0}{0}{1}
		\fullvertex{$\Gamma$}{0}{0}{1}
		\abubblefullfull{0.3}{0}{1}{1}{1.2}
		\fullvertexwide{$\dot{\gamma}_{\bar{a}}^{(\ell+1)}$}{1.86}{0}{1.2}{0.3}
		\arrowsrighthalffull{2.16}{0}{1.2}
		\tikzunderbrace{$\dot{\gamma}_{a,R}^{(\ell+2)}$}{-0.6}{2.82}{-1}{-0.6}
	}
	\,,
	\label{eq:mfRG:mfRG:Flow_equations:3l}
\end{align}
\end{subequations}
\end{widetext}
and analogously in the other two channels $p$ and $t$. Here, $\gamma_{\bar{r}}^{(\ell)} = \sum_{r'\neq\bar{r}} \gamma_{r'}^{(\ell)}$ and the last equation \eqref{eq:mfRG:mfRG:Flow_equations:3l} applies for all higher loop orders $\ell+2 \geq 3$. The double dashed bubble in Eq.\,\eqref{eq:mfRG:mfRG:Flow_equations:1l} corresponds to a sum of two terms, $\dot{\Pi} = \dot{G}G + G\dot{G}$, where $\dot{G} = S + G\dot{\Sigma}G$ with the single-scale propagator $S = \left.\partial_\Lambda G\right|_{\Sigma=\text{const.}}$ and the Katanin substitution\cite{katanin_fulfillment_2004}.

The multiloop flow equation for the self-energy reads 
\begin{subequations}
\begin{align}
	\tikzm{mfRG-ml-dSigma}{
		\selfenergywithlegs{$\dot\Sigma$}{0}{0}{1.2}
	}
	\
	&= \
	- \
	\tikzm{mfRG-ml-dSigma_std}{
		\fullvertex{$\Gamma$}{0}{0}{1.2}
		\arrowslowerhalffull{0}{0}{1.2}
		\loopfullvertex{lineWithArrowCenterCenter}{0}{0}{1.2}
		\draw[linePlain] (0,0.76) -- (0,1.16);
	} \label{eq:SE-one-loop}
	\\
	& \quad \ \ \underbrace{
		- \
		\tikzm{mfRG-ml-dSigma_tbar}{
			\fullvertex{$\dot\gamma_{\bar{t},C}$}{0}{0}{1.2}
			\arrowslowerhalffull{0}{0}{1.2}
			\loopfullvertex{lineWithArrowCenterCenter}{0}{0}{1.2}
			\node at (0,1.01) {};
		}
	}_{\dot\Sigma_{\bar{t}}}
	\
	\underbrace{
		- \
		\tikzm{mfRG-ml-dSigma_t}{
			\fullvertex{$\Gamma$}{0}{0}{1.2}
			\arrowslowerhalffull{0}{0}{1.2}
			\selfenergy{$\dot\Sigma_{\bar{t}}$}{0}{1}{1.2}
			\draw[lineWithArrowCenter] (0.36,0.36) to [out=45, in=270] (0.55,0.65) to [out=90, in=0] (0.36,1);
			\draw[lineWithArrowCenter] (-0.36,1) to [out=180, in=90] (-0.55,0.65) to [out=270, in=135] (-0.36,0.36);
		}
	}_{\dot\Sigma_t}
	\,,
	\label{eq:mfRG:mfRG:Selfenergy_flow}
\end{align}
\end{subequations}
with $\dot{\gamma}_{\bar{t},C} = \sum_{\ell} (\dot{\gamma}_{a,C}^{(\ell)} + \dot{\gamma}_{p,C}^{(\ell)})$, where the single-dashed line denotes the single-scale propagator $S$ from above.

Historically, fRG flow equations have been derived from a generating functional, yielding an exact hierarchy of flow equations, coupling $n$-point vertices of increasing order\cite{wetterich_exact_1993}. As the six-point vertex, which contributes to the flow equation of $\Gamma$, see Eqs.\,(19) in Ref.\,\onlinecite{main_paper}, is inaccessible numerically, its contribution often is neglected completely, resulting in the so-called ``one-loop'' flow equations. This, however, results in an unphysical dependence of the final result of the flow on the choice of regulator (as the flow equations no longer constitute total derivatives), and also introduces a bias towards ladder diagrams \cite{kugler_multiloop_2018, Kugler_2018}.

The multiloop framework builds upon the one-loop scheme by iteratively adding precisely those two-particle reducible diagrammatic contributions to the flow equations which are required to reinstate total derivatives w.r.t.~$\Lambda$ and thereby makes it reproduce the solution of the parquet equations. In that sense, it provides an alternative scheme for solving the parquet iterations via differential equations. From a computational standpoint, the mfRG flow equations introduce a complication compared to the one-loop flow equations, in that the right-hand sides of the flow equations for both $\Gamma$ and $\Sigma$ involve the differentiated self-energy and vertex. In order to still be able to use standard algorithms for ordinary differential equations, a scheme was outlined in Ref.\,\onlinecite{kugler_multiloop_2018} to include those differentiated quantities iteratively: Starting from the one-loop term Eq.\,\eqref{eq:SE-one-loop} to evaluate the flow equations \eqref{eq:mfRG:mfRG:Flow_equations} for $\Gamma$, these are then iterated with the multiloop corrections \eqref{eq:mfRG:mfRG:Selfenergy_flow} at every step of the flow until convergence is reached. The number of iterations required for convergence at this point can again be reduced using Anderson acceleration, as described in Sec.\,\ref{sec:parquet}. 

From the code, an mfRG-flow computation can be started with the function \code{n\_loop\_flow}, which requires only the string for the name of the output file and a set of parameters. It is overloaded to enable checkpointing, i.e. it is possible to continue a previously started computation from a given iteration. This is particularly useful for demanding jobs that take a long time and is highly recommended to any user. 

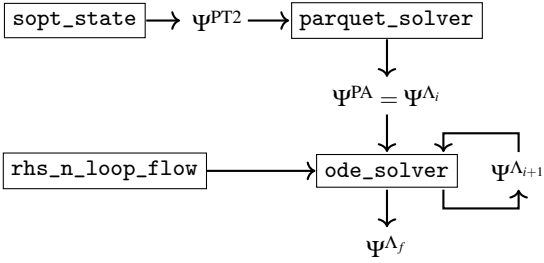
\begin{figure}
    \begin{tikzpicture}
        \node[draw] at (0.1,0) {\code{sopt\_state}};
        \node at (2, 0) {$\Psi^{\text{PT2}}$};
        \node[draw] at (4.25, 0) {\code{parquet\_solver}};
        \node at (4.25, -1) {$\Psi^{\text{PA}} = \Psi^{\Lambda_i}$};
        \node[draw] at (4.25, -2) {\code{ode\_solver}};
        \node at (6, -2) {$\Psi^{\Lambda_{i+1}}$};
        \node at (4.25, -3) {$\Psi^{\Lambda_f}$};
        \node[draw] at (0.5, -2) {\code{rhs\_n\_loop\_flow}};
        \draw[->, thick] (1.05, 0) -- (1.5, 0);
        \draw[->, thick] (2.4, 0) -- (2.95, 0);
        \draw[->, thick] (4.25, -0.25) -- (4.25, -0.75);
        \draw[->, thick] (4.25, -1.25) -- (4.25, -1.75);
        \draw[->, thick] (4.25, -2.25) -- (4.25, -2.75);
        \draw[->, thick] (1.85, -2) -- (3.3, -2);
        \draw[->, thick] (5, -2.25) -- (5, -2.5) -- (6, -2.5) -- (6, -2.25);
        \draw[->, thick] (6, -1.75) -- (6, -1.5) -- (5, -1.5) -- (5, -1.75);
    \end{tikzpicture}
    \caption{Schematic depiction of the function \code{n\_loop\_flow}.}
    \label{fig:flow}
\end{figure}

The function \code{n\_loop\_flow} works as shown in Fig.~\ref{fig:flow}.
It first initializes a state using PT2 with the function \code{sopt\_state}, see Sec.\,\ref{sec:PT2}, and then uses this result as a seed for a full parquet computation at the initial value of the regulator $\Lambda_i$ with the \code{parquet\_solver} function, see Sec.\,\ref{sec:parquet}. This provides a suitable starting point for the following mfRG calculation.

\begin{figure}
    \begin{tikzpicture}
        \node at (0,0) {\boxed{\code{selfEnergyOneLoopFlow}}};
        \node at (0,-1) {$\dot{\Sigma}^{\ell=1}$};
        \node at (0,-2) {\boxed{\code{vertexOneLoopFlow}}};
        \node at (0,-3) {$\dot{\Gamma}^{\ell=1}$};
        \node at (0,-4) {\boxed{\code{calculate\_dGammaL}}};
        \node at (0,-4.55) {\boxed{\code{calculate\_dGammaR}}};
        \node at (0,-5.5) {$\dot{\Gamma}^{\ell=2}$};
        \node at (0,-6.5) {\boxed{\code{calculate\_dGammaL}}};
        \node at (0,-7.05) {\boxed{\code{calculate\_dGammaC}}};
        \node at (0,-7.6) {\boxed{\code{calculate\_dGammaR}}};
        \node at (0,-8.5) {$\dot{\Gamma}^{\ell=n}$};
        \node at (0,-9.5) {\boxed{\code{selfEnergyFlowCorrections}}};
        \node at (3.5,-9.5) {$\dot{\Sigma}^{\ell=n}$};
        \draw [thick, decorate, decoration = {calligraphic brace, raise=5pt, amplitude=5pt}] (-1.5,-7.05 - 0.8) --  (-1.5,-7.05 + 0.8);
        \draw[->, thick] (-2, -7.05) -- (-2.3, -7.05) -- (-2.3, -6) -- (-1, -6) -- (-1, -6.25);
        \node[align=left] at (-2.3, -5.8) {\small iterate up to $\ell=n$};
        \draw[->, thick] (0, -0.25) -- (0, -0.7);
        \draw[->, thick] (0, -1.3) -- (0, -1.75);
        \draw[->, thick] (0, -2.25) -- (0, -2.7);
        \draw[->, thick] (0, -3.3) -- (0, -3.75);
        \draw[->, thick] (0, -4.8) -- (0, -5.2);
        \draw[->, thick] (0, -5.8) -- (0, -6.25);
        \draw[->, thick] (0, -7.85) -- (0, -8.2);
        \draw[->, thick] (0, -8.8) -- (0, -9.25);
        \draw[->, thick] (2.2, -9.5) -- (3, -9.5);
        \draw[->, thick] (3.5, -9.2) -- (3.5, -2) -- (1.55, -2);
        \node[align=left] at (3.5, -1.8) {\small iterate until convergence};
        \draw[->, thick] (-0.5, -3) -- (-2.3, -3) -- (-2.3, -5.2) -- (-0.5, -5.2) -- (-0.5, -6.25);
    \end{tikzpicture}
    \caption{Structure of the function \code{rhs\_n\_loop\_flow}, including multiloop iterations up to loop order $\ell=n$ and self-consistent self-energy iterations due to the multiloop corrections.}
    \label{fig:rhs}
\end{figure}
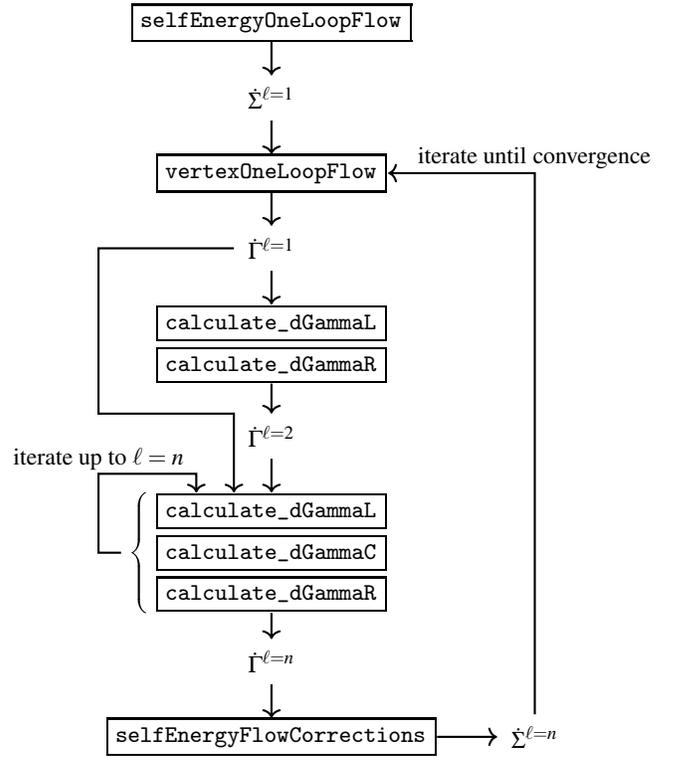

The \code{ode\_solver} function carries out the actual calculation of solving the mfRG flow. It uses an instance of the \code{rhs\_n\_loop\_flow\_t} class, which provides a wrapper to the function \code{rhs\_n\_loop\_flow}, which in turn evaluates the right-hand side of the flow equations given an input state at a given value of $\Lambda$. This is done iteratively by loop order according to the flow equations \eqref{eq:mfRG:mfRG:Flow_equations}, including self-consistent iterations for the self-energy starting at 3-loop level, as outlined above. The function \code{rhs\_n\_loop\_flow} is structured as shown in Fig.~\ref{fig:rhs}.
Starting from the self-energy and vertex from the previous step of the ODE-solver, it evaluates the right-hand sides of the flow equations by first computing the one-loop term of the flow equation for the self-energy, Eq.\,\eqref{eq:SE-one-loop}, with the function \code{selfEnergyOneLoopFlow}. The result is then used to evaluate the one-loop term of the flow equation for the two-particle reducible vertices $\dot{\gamma}_r$, Eq.\,\eqref{eq:mfRG:mfRG:Flow_equations:1l}, involving a fully differentiated bubble. 
Then, the one-loop result is used to evaluate the two-loop contribution, Eq.\,\eqref{eq:mfRG:mfRG:Flow_equations:2l}, which consists of two terms: One where the differentiated one-loop contribution $\dot{\gamma}_r^{(1)}$ is used as the left part of a bubble contraction with the full vertex and one where it is used on the right side. These two terms are computed using the functions \code{calculate\_dGammaL} and \code{calculate\_dGammaR}, respectively. Next, the three-loop contribution is computed, which involves both the one-loop and the two-loop result, see Eq.\,\eqref{eq:mfRG:mfRG:Flow_equations:3l}. Again, the functions \code{calculate\_dGammaL} and \code{calculate\_dGammaR} are invoked and in addition the function \code{calculate\_dGammaC} to compute the ``center term'' involving two bubble contractions of $\dot{\gamma}_r^{(1)}$ with the full vertex; once to the left and once to the right. As the structure of the flow equations does not change from this point on, this part is iterated until the maximally desired loop number $n$ (which is given as a runtime parameter, see Sec.\,\ref{sec:parameters}) is reached. The resulting center terms of the $a$ and $p$ channels are then used to evaluate the multiloop corrections to the self-energy, according to Eq.\,\eqref{eq:mfRG:mfRG:Selfenergy_flow}. This updates the differentiated bubble used in the computation of the one-loop terms $\dot{\gamma}_r^{(1)}$, such that the whole process is finally iterated from that point on, until convergence is reached, as determined by the parameters \code{tol\_selfenergy\_correction\_abs} and \code{tol\_selfenergy\_correction\_rel}, see Sec.\,\ref{sec:parameters}. All functions invoked by \code{rhs\_n\_loop\_flow} of course make heavy use of the main functionality outlined in Sec.\,\ref{sec:functionality}.

As a side note, it is possible to parametrize the vertex using the single-boson exchange (SBE) decomposition \cite{PhysRevB.99.235106, PhysRevB.100.155149, PhysRevB.100.245147, PhysRevB.102.235133, PhysRevB.102.195131, PhysRevResearch.3.013149} and to rewrite the mfRG flow equations in this language, as outlined in Ref.\,\onlinecite{Gievers2022}. This is achieved by setting the flag \code{SBE\_DECOMPOSITION} to $1$. Two versions of the SBE approximation can be used, known as ``SBEa'' and ``SBEb'' in the literature \cite{Fraboulet2022}. Which version is to be used is controlled by the flag \code{USE\_SBEb\_MFRG\_EQS}, see Sec.\,\ref{sec:parameters}. This functionality is so far, however, only implemented in the MF. We therefore refrain from providing further details here.

In the final two parts of this section, we discuss the ODE-solver and the different flow schemes.

\subsubsection{Details on the ODE-solver}\label{sec:ODE-details}

To solve the mfRG flow equations accurately, a Cash-Karp routine \cite{10.1145/79505.79507} is implemented, which constitutes a fourth-order Runge-Kutta solver with adaptive step size control. An adaptive step-size control is crucial for obtaining accurate results and is hereby strongly recommended for solving fRG flow equations precisely. For a good first guess of the step size in the $\Delta$-flow (see Par.\,\ref{sec:delta-flow}), the flow parameter is reparametrized as $\Lambda(t) = 5 t |t|/\sqrt{1 - t^2}.$ For equidistant $t$, this parametrization provides large steps for large $\Lambda$ and small steps for small $\Lambda$. This is sensible in the context of the $\Delta$-flow, where $\Lambda$ is gradually reduced to enter into ever more challenging parameter regimes.

\subsubsection{Flow schemes}\label{sec:flow_schemes}

In fRG, one chooses a regulator introduced into the bare propagator $G_0 \rightarrow G_0^\Lambda$, i.e.\, the flow scheme. While the solution of a truncated set of fRG flow equations will depend on this choice, a converged multiloop flow will not, as it reproduces the self-consistent solution of the parquet equations. It is generally advisable to choose the most convenient flow scheme for the problem at hand. In particular, the fRG flow can be used to compute a full parameter sweep in one go, by choosing a physical parameter as the regulator. Compared to direct solutions of the parquet equations, which have to be computed individually at every point in parameter space, this makes mfRG computations more economical, provided they can be quickly converged in the loop order. In the following, we outline the flow schemes that have been implemented and can be used by setting the \code{REG} flag and the \code{Lambda\_ini} and \code{Lambda\_fin} parameters accordingly, see Tables \ref{tab:preprocessor_macros} and \ref{tab:global_parameters}.

\paragraph{$\Delta$-flow}\label{sec:delta-flow}

The hybridization flow\cite{jakobs_functional_2010} uses $\Delta$ as the flow parameter, starting at a very large value and decreasing $\Delta$ to a smaller value, keeping the other parameters $U$ and $T$ fixed. The hybridization flow thus performs a parameter sweep in $U/\Delta$ for fixed $T/U$. The Keldysh fRG single-scale propagator reads
\begin{align*}
    S^R(\nu) = \partial_\Delta \left.G^R(\nu)\right|_{\Sigma = \mathrm{const.}} = -i[G^R(\nu)]^2.
\end{align*}
In practice, we start the fRG flow from a solution of the parquet equations at large $\Delta$ (small $U/\Delta$), where that solution can be easily obtained. For historical reasons, the hybridization flow is implemented as 
\begin{equation}
    G^R_\Lambda(\nu) = \frac{1}{\nu - \varepsilon_d + i (\Gamma + \Lambda)/2 - \Sigma^R_\Lambda(\nu)}
\end{equation}
inside the code, where $\Gamma$ is fixed to some arbitrary value and $\Lambda$ is used to fix the hybridization $\Delta = (\Gamma + \Lambda)/2$.
Note that keeping $T/U$ fixed during the $\Delta$-flow is a somewhat unconventional choice, as in most works on the AM the scale $T/\Delta$ is kept constant. As explained in Ref.\,\onlinecite{walter_keldyshfrg_2022}, keeping $T/\Delta$ fixed during the $\Delta$-flow would lead to additional sharply peaked terms in the single-scale propagator and has hence not been pursued yet.

\paragraph{$U$-flow}

An alternative to the $\Delta$-flow is the following flow scheme first introduced in Ref.\,\onlinecite{interaction-flow},
\begin{align}
    G^R_\Lambda(\nu) = \frac{\Lambda}{\nu - \varepsilon_d + i \Delta - \Lambda \, \Sigma^R_\Lambda(\nu)},
\end{align}
starting at $\Lambda_i = 0$ (or very small, in practice) and flowing towards $\Lambda_f = 1$. The corresponding single-scale propagator then reads
\begin{align}
    S^R(\nu) &= \partial_\Lambda \left.G^R(\nu)\right|_{\Sigma = \mathrm{const.}} = \frac{\nu - \varepsilon_d + i \Delta}{[\nu - \varepsilon_d + i \Delta - \Lambda \, \Sigma^R_\Lambda(\nu)]^2}.
\end{align}
This flow scheme is called interaction- or $U$-flow because increasing $\Lambda$ effectively amounts to increasing $U$. This can be shown by a simple rescaling argument: A bare diagram for $\Sigma$ (or $\Gamma$) at order $n$ has $n$ factors of $U$ and $2n-1$ (or $2n-2$) factors of $G_{0, \Lambda}$, each contributing one factor of $\Lambda$. The same scaling behaviour in $\Lambda$ can be achieved without a $\Lambda$-dependent $G_0$ by multiplying $U$ with $\Lambda^2$ and dividing out an extra $\Lambda$ (or $\Lambda^2$. It hence holds that
\begin{subequations}
\begin{align}
    \Sigma_\Lambda(U) & = \Sigma(\Lambda^2 U)/\Lambda \\
    \Gamma_\Lambda(U) &= \Gamma(\Lambda^2 U)/\Lambda^2.
\end{align}
\end{subequations}
Note that at zero temperature, the two flow schemes discussed so far should be equivalent: For $T=0$, the only energy scales of the AM in the wideband limit are $U$ and $\Delta$, so there is only one external parameter $U/\Delta$ and it does not matter whether $U$ is increased or $\Delta$ is decreased.\\
Historically, the $U$-flow has not been very popular, as it does not regulate IR divergences\cite{RevModPhys.84.299}. Nevertheless, it can be used for the AM. In Ref.\,\onlinecite{main_paper} we found that, for a truncated 1-loop Keldysh fRG flow at \emph{finite} $T$, this scheme produces inferior results compared to the $\Delta$-flow when benchmarked against numerically exact NRG data. Still, the $U$-flow has the nice property that it keeps $T/\Delta$ fixed.

\paragraph{$T$-flow}

Using temperature as the fRG flow parameter has been popular in the past when performing fRG computations in the MF\cite{honerkamp_t-flow, schneider2023temperature}. It has been argued that temperature cannot be used for this purpose in Keldysh fRG computations\cite{jakobs_functional_2010}, the reason being that a truncated fRG flow does not preserve fluctuation-dissipation relations (FDRs). However, solutions of the parquet equations \emph{do} fulfill the FDRs. If the FDRs are not used explicitly during mfRG calculations (as this would mix FDRs at different temperatures and hence introduce an inconsistency) it should be possible to obtain these solutions also by converging an mfRG flow.
Instead of the standard FDR, which relates $G^K$ and $G^R$, in this scheme, the general expression for the Keldysh component of the propagator should be used, which reads\cite{walter_keldyshfrg_2022}
\begin{align}
    G^K(\nu) &= G^R(\nu)\left[\Sigma^K(\nu) -2i\Delta\tanh\left(\tfrac{\nu}{2T}\right)\right] G^A(\nu).
\end{align}
The Keldysh component of the single-scale propagator is then
\begin{align}
    S^K(\nu) = \left. \partial_T G^K(\nu)\right|_{\Sigma = \mathrm{const.}} &= \frac{i\Delta\nu}{T^2\cosh^2\left(\tfrac{\nu}{2T}\right)} |G^R(\nu)|^2.
\end{align}
Note that its retarded component is zero, $S^R(\nu)=0$, as $G^R(\nu)$ does not depend explicitly on $T$.
While preliminary numerical results suggest that this scheme indeed performs well, a systematic study of the temperature flow in Keldysh fRG is left for future work. So far, the time of writing, the temperature flow described above can only be used in the KF; corresponding regulators in the MF as in Refs.\,\onlinecite{honerkamp_t-flow} and \onlinecite{schneider2023temperature} have not been implemented.

\paragraph{$\nu$-flow}

Using a frequency regulator of the form $G_{0,\Lambda}(i\nu) = G_0(i\nu) \Theta_\Lambda(i\nu)$ with $\Theta_\Lambda(i\nu) = \nu^2 / (\nu^2 + \Lambda^2)$ has been a popular choice in the literature for (m)fRG calculations in the Matsubara formalism \cite{10.21468/SciPostPhys.6.1.009, PhysRevResearch.4.023050}. However, in this form the frequency regulator cannot be used in the Keldysh formalism, as analytical continuation of $\Theta_\Lambda(i\nu)$ gives $\Theta^R_\Lambda(\nu) = \nu^2 / (\nu^2 - \Lambda^2 + 2|\nu|i0^+)$ with a branch cut for $\nu<0$. One can, however, change the form of the regulator to $\Theta_\Lambda(i\nu) = |\nu|/(|\nu| + \Lambda)$, for which the retarded counterpart reads
\begin{align}
    \Theta^R_\Lambda(\nu) = \frac{\nu}{\nu + i\Lambda},
\end{align}
which \emph{is} a well-behaved function. This choice is implemented as
\begin{align}
    G^R_\Lambda(\nu) &= \frac{\Theta^R_\Lambda(\nu)}{\nu - \varepsilon_d + i \Delta - \Theta^R_\Lambda(\nu) \Sigma^R_\Lambda(\nu)}.
\end{align}
The corresponding single-scale propagator then reads
\begin{align}
    S^R(\nu) &= -\frac{i}{\nu} \frac{[\Theta^R_\Lambda(\nu)]^2(\nu - \varepsilon_d + i \Delta)}{[\nu - \varepsilon_d + i \Delta - \Theta^R_\Lambda(\nu) \Sigma^R_\Lambda(\nu)]^2}.
\end{align}
With this choice, all causality relations and FDRs are satisfied. However, this regulator has two drawbacks compared to the other flow schemes: First, it does not produce a parameter sweep, as $\Lambda$ does not directly correspond to a physical parameter. Second, computations become ever more challenging for smaller $\Lambda$: Even if all correlation functions are reasonably smooth in frequency space for $\Lambda=0$, for small but finite $\Lambda$, they exhibit sharp features. While this is not an issue for finite-temperature Matsubara calculations, where only sums over discrete Matsubara frequencies are performed, it turns out to be a major inconvenience in the Keldysh context.

\section{Conclusion}\label{sec:conclusion}

In this paper, we outlined the structure and the design of our C++ codebase for diagrammatic calculations of the AM in the Keldysh formalism. We explained the building blocks for representing real-frequency correlation functions and the central routines used to compute them. We elaborated on all performance-critical aspects, allowing one to handle the three-dimensional frequency dependence of the four-point vertex and summarized the implementation of the parquet and mfRG equations. By discussing the most convenient features of the codebase -- modularity, flexibility, performance, and scalability -- but also some of its design flaws in  detail, we hope to provide guidance and inspiration to others who plan to write a code for similar purposes.

Our codebase forms the basis for numerous future projects involving dynamical correlation functions of electronic many-body systems. Since the AM is very well understood, we want to generalize our treatment to more complicated models with unexplored physics like lattice models, possibly including multiple bands. The main problem in that regard is the numerical complexity: In addition to their real-frequency Keldysh structure, all functions would acquire momentum dependencies and orbital indices. Parametrizing those naively appears prohibitively costly. Fortunately, the new quantics tensor cross interpolation (QTCI) scheme \cite{Shinaoka2022, nunez_fernandez_learning_2022, ritter_quantics_2023} is currently being developed, which can be used to obtain highly compressed tensor network representations of correlation functions and promises exponential reductions in computational costs. It is remains to be seen how efficiently the Keldysh four-point vertex can be compressed using this method. If it turned out to be highly compressible, one could combine the diagrammatic approaches outlined here with non-perturbative results from dynamical mean-field theory to access truly strongly correlated parameter regimes (see related works \cite{PhysRevB.99.041115, PhysRevB.99.104501, PhysRevResearch.4.013034} in the MF). 
In particular, computing non-local real-frequency dynamical vertex corrections beyond DMFT to observables like optical conductivities with high precision is a formidable long-term goal.

Another possible future direction relates to nonequilibrium phenomena, for example the influence of the full four-point vertex on observables like differential conductivities\cite{jakobs_nonequilibrium_2010, PhysRevB.68.155310}. Nonequilibrium physics has been the most popular application of the KF in the past and the AM with a finite bias voltage is tractable with only minor increase in both the numerical costs and the implementation effort. 

In order to leverage ongoing efforts in the QTCI framework, an interface to the corresponding Julia package \cite{QuanticsTCI.jl} would be required. Given that, in recent years, multiple Julia codes have been developed to perform calculations of two-particle correlation functions \cite{PhysRevResearch.4.023185, 10.21468/SciPostPhysCodeb.24, 10.21468/SciPostPhysCodeb.24-r0.1, nils_niggemann_2023_10255230}, it would be natural to switch to that language in the future, especially since it allows much simpler structures and in general perform almost as well as C++.

\begin{acknowledgments}
 We thank Marc Ritter, Marcel Gievers, Benedikt Schneider, Björn Sbierski, Nils Niggemann, Dominik Kiese, Aiman Al-Eryani, Lennart Klebl and Jacob Beyer for insightful conversations about fRG code structures over the years and Markus Frankenbach for comments on the manuscript.
 
 NR acknowledges funding from a graduate scholarship from the German Academic Scholarship Foundation (``Studienstiftung des deutschen Volkes'') and additional support from the ``Marianne-Plehn-Programm'' of the state of Bavaria. AG and JvD were supported by the Deutsche Forschungsgemeinschaft under Germany’s Excellence Strategy EXC-2111 (Project No.~390814868), and the Munich Quantum Valley, supported by the Bavarian state government with funds from the Hightech Agenda Bayern Plus. FBK acknowledges support from the Alexander von Humboldt Foundation through the Feodor Lynen Fellowship. The Flatiron Institute is a division of the Simons Foundation.  
\end{acknowledgments}
\section*{Conflict of Interest Statement}
The authors have no conflicts to disclose.

\section*{Code availability statement}

The code, which includes a link to the technical documentation, can be found online, see Ref.\,\onlinecite{Ritz_KeldyshDiagrammatics_A_C}. It is published under the MIT license.

\section*{References}
\bibliography{bibliography}

\end{document}